\documentclass[prc,preprint,floatfix,amsfonts,tightenlines, nofootinbib,superscriptaddress]{revtex4-1}

\pdfoutput=1

\usepackage{graphicx}  %
\usepackage{bm}  %
\usepackage{amssymb}
\usepackage{dcolumn}

\usepackage[normalem]{ulem}
\usepackage{amsmath,simplewick}
\usepackage{subfigure}

\usepackage[latin1]{inputenc}
\usepackage[english]{babel}
\usepackage[T1]{fontenc}
\usepackage{amsmath}
\usepackage{amsfonts}
\usepackage{geometry}
\usepackage{bbm}
\usepackage{float}
\usepackage{dsfont}

\newcommand{\dif}[1]{\ensuremath{\medspace \mbox{d} #1}}

\newcommand{\p}[1]{\ensuremath{\mathbf{#1}}}

\newcommand{\bra}[1]{\ensuremath{\langle #1 |}}
\newcommand{\ket}[1]{\ensuremath{| #1 \rangle}}

\newcommand{\Li}{\ensuremath{\,\mbox{Li}}}

\newcommand{\Hinf}{H_{\infty}}
\newcommand{\Vinf}{V_{\infty}}
\newcommand{\psilamalpha}{\psi^{\Lambda}_\alpha}
\newcommand{\psilamalphaA}{\psi^{\Lambda}_{\alpha_{,A}}}

\newcommand{\psilamalphastar}{\psi^{\Lambda *}_\alpha}

\newcommand{\psiinfalpha}{\psi^{\infty}_{\alpha}}
\newcommand{\psiinfalphaA}{\psi^{\infty}_{\alpha_{,A}}}

\newcommand{\psiinfalphastar}{\psi^{\infty *}_{\alpha}}

\graphicspath{{./PicsData/}{./PicsData1D/}}

\begin{document}

%

\newcommand{\beqn}{\begin{equation}}
\newcommand{\eeqn}{\end{equation}}
\newcommand{\bea}{\begin{eqnarray}}
\newcommand{\eea}{\end{eqnarray}}

\newcommand{\vlowk}{V_{{\rm low}\,k}}
\newcommand{\vsrg}{V_s}
\newcommand{\tlowk}{T_{{\rm low}\,k}}
\newcommand{\trel}{T_{\rm rel}}
\newcommand{\vnn}{V_{\rm NN}}

\newcommand{\fm}{\, \text{fm}}
\newcommand{\fmi}{\, \text{fm}^{-1}}
\newcommand{\mev}{\, \text{MeV}}

\newcommand{\la}{\langle}
\newcommand{\ra}{\rangle}
\newcommand{\ts}{\textstyle}
\newcommand{\wt}{\tilde}
\newcommand{\wh}{\widehat}

\newcommand{\qvec}{{\bf q}}
\newcommand{\kvec}{{\bf k}}
\newcommand{\kpvec}{{\bf k'}}
\newcommand{\kmax}{k_{\rm max}}

\newcommand{\adag}{a^\dagger}
\newcommand{\adaggera}{a^\dagger_{q} a^{{\protect\phantom\dagger}}_{q}}
\newcommand{\adaggeraop}{(\adaggera)_s}
\newcommand{\Hzero}{H^{\rm bd}}
\newcommand{\flow}{s}
\newcommand{\oneoverr}{r^{-1}}

\newcommand{\mystrut}{\rule[-1mm]{0mm}{5.5mm}}

\newcommand{\Oop}{\wh{O}}
\newcommand{\projP}{\mathcal{P}}
\newcommand{\projQ}{\mathcal{Q}}

%
%
%
%


%
\title{High-momentum tails from low-momentum effective theories}

\author{S.K.\ Bogner}\email{bogner@nscl.msu.edu}
\affiliation{National Superconducting Cyclotron Laboratory and Department of Physics and Astronomy, Michigan State University, East Lansing, MI 48824}
\author{D.\ Roscher}\email{dietrich.roscher@uni-jena.de}
\affiliation{National Superconducting Cyclotron Laboratory and Department of Physics and Astronomy, Michigan State University, East Lansing, MI 48824}
\affiliation{Theoretisch-Physikalisches Institut, Friedrich-Schiller-Universit\"at Jena, Max-Wien-Platz 1, D-07743 Jena, Germany}
\date{\today}

\begin{abstract}
In a recent work~\cite{Anderson:2010aq}, Anderson \emph{et al.} used the renormalization group (RG) evolution of the momentum distribution to show that, under appropriate conditions, operator expectation values exhibit factorization in the two-nucleon system. Factorization is useful because it provides a clean separation of long- and short-distance physics, and suggests a possible interpretation of the universal high-momentum dependence and scaling behavior found in nuclear momentum distributions. In the present work, we use simple decoupling and scale-separation arguments to extend the results of Ref.~\cite{Anderson:2010aq} to arbitrary low-energy $A$-body states. Using methods that are reminiscent of the operator product expansion (OPE) in quantum field theory, we find that the high-momentum tails of momentum distributions and static structure factors factorize into the product of a universal function of momentum that is fixed by two-body physics, and a state-dependent matrix element that is the same for both and is sensitive only to low-momentum structure of the many-body state. As a check, we apply our factorization relations to two well-studied systems, the unitary Fermi gas and the electron gas, and reproduce known expressions for the high-momentum tails of each. 
\end{abstract}

\maketitle

\section{Introduction}
\label{sec:intro}

Renormalization group (RG) methods play an important role in {\it ab initio} nuclear theory by extending the range of many computational methods and improving their convergence patterns \cite{Bogner:2003wn, Bogner:2009bt, Furnstahl:2012fn}. There are numerous RG methods that have been successfully applied to nuclear few- and many-body systems in recent years~\cite{Bogner:2009bt}. While the details differ, all such methods decouple low- and high-momentum degrees of freedom in a manner that leaves low-energy observables invariant. 
In this paper, we will denote the momentum scale at which this decoupling occurs by $\Lambda$.
In methods such as the Lee-Suzuki-Okubo similarity transformation method or the related $\vlowk$ approach, $\Lambda$ is a floating cutoff beyond which high-momentum states have been integrated out~\cite{Epelbaum:1998na, Bogner:2001gq,Bogner:2006vp}.  For other methods, such as the similarity renormalization group (SRG) approach, $\Lambda$ gives a measure of how band-diagonal the Hamiltonian is in momentum space~\cite{Bogner:2006pc}.  In all cases, $\Lambda$ serves as a ``resolution scale''  since dynamics above and below this scale are effectively decoupled~\cite{Bogner:2009bt,Jurgenson:2007td}.  

We emphasize that while
observable quantities (such as cross sections) do not change, the physics interpretation can (and generally
does) change with resolution. It is a common misconception that at low-resolution,  one is unable to describe phenomena that, at high-resolution, are associated with the high-momentum components of low-energy wave functions. A prototypical example is the $(e,e'p)$ process at large momentum
transfers, where theoretical analyses relate such
experiments to nuclear momentum distributions if
the impulse approximation is assumed valid for a high-cutoff interaction \cite{Frankfurt:2008zv}.
Calculations find nearly universal scaling of the high-momentum tails, which is 
interpreted in terms of short-range correlations in the nuclear wave functions~\cite{Pieper:1992gr}. Naively it might be thought that this physics is beyond the reach
of low-momentum approaches, for which wave functions have drastically
reduced short-range correlations.  However, this is not the case:  the experimental cross section is unchanged if the corresponding operator is consistently evolved under the RG, even if the evolved wave function has almost no high-momentum strength.  The formal relationship between an operator $\hat{O}^{\Lambda_0}$ at an initial high-resolution scale $\Lambda_0$, and the consistently-evolved effective operator  $\hat{O}^{\Lambda}$ at the low-resolution scale $\Lambda$ is \emph{defined} by 
\begin{equation}
\label{eq:Oeff}
\bra{\psi_n^{\Lambda_0}}\hat{O}^{\Lambda_0}\ket{\psi_n^{\Lambda_0}} = \bra{\psi_n^{\Lambda}}\hat{O}^{\Lambda}\ket{\psi_n^{\Lambda}}\,.
\end{equation}

In the $(e,e'p)$ example, one might worry that the consistent evolution embodied by Eq.~\ref{eq:Oeff} is computationally intractable because the evolved momentum occupation operator might be too
complicated in practice (e.g., strong non-localities and sizable many-body components, etc.).  In Ref.~\cite{Anderson:2010aq}, some of these questions were addressed by examining the consistent SRG evolution of various operators, including momentum distributions and electromagnetic form factors in the deuteron. There, all operators were found to flow to smooth, low-momentum forms, exhibiting many of the same simplifications as RG-evolved interactions. More interestingly, under certain kinematic conditions it was found that operator expectation values exhibit \emph{factorization}, which provides a clean separation of long- and short-distance physics and an alternative interpretation of the universal high-momentum dependence and scaling behavior~\cite{Anderson:2010aq}. 

The proof of factorization presented here and in Ref.~\cite{Anderson:2010aq} follows straightforwardly from decoupling and the separation of scales, and is reminiscent of the operator product expansion (OPE) in quantum field theory. The OPE was developed for the evaluation of singular products of local field operators at small separation~\cite{Wilson:1969zs,Wilson:1972ee}. The utility of the OPE rests on factorization; short-distance details decouple from long-distance dynamics. Factorization enables one, for example, to separate the momentum and distance scales in hard-scattering processes in terms of perturbative QCD and parton distribution functions. While the methods used in the present paper share several similarities with the OPE, a precise connection has not yet been made. One key difference is that, in the framework of a local quantum field theory, the OPE gives a controlled expansion since the dependence of the Wilson coefficients on the separation $\p{r}$ is fixed by the scaling dimensions of the corresponding local operators.  In the present paper, however, we work in the general domain of non-relativistic quantum mechanics (i.e., no assumption of a local QFT). Therefore, we cannot make precise statements about the scaling behavior of terms when we expand Fock space operators at one resolution scale $\Lambda$ in terms of the corresponding operators at another scale $\Lambda_0 \ge \Lambda$. While the factorization formulas in the present context are not as controlled as those derived in a local quantum field theory using the OPE, they nevertheless provide tools that let us parameterize the high-momentum components of operators which would normally require degrees of freedom we do not retain. We can, for example, build effective few-body operators containing state-independent functions of high momenta that can be measured directly in few-body experiments. These operators can then be employed to make predictions for $A$-body systems. 

In this paper, we generalize previous developments~\cite{Anderson:2010aq} to derive scaling relations for the high-momentum tails of momentum distributions and static structure factors in arbitrary low-energy $A$-body states. In both instances, we find that the expectation value of the corresponding operator factorizes into the product of a universal function associated with high-momentum (short-distance) physics, and a state-dependent number associated with low-momentum (long-distance) structure. The outline of the rest of this paper is as follows: In Section~\ref{sec:twobody}, we review the proof from Ref.~\cite{Anderson:2010aq} that expectation values of high-momentum probes factorize in the $A=2$ system. In Section~\ref{sec:Abodyfac}, we recast the discussion of factorization in a second-quantized language and use it to derive universal scaling relations for momentum distributions and static structure factors in general $A$-body systems. As a test of these relations, in Section~\ref{sec:examples} we apply them to two well-studied many-body systems - the unitary Fermi gas and the electron gas - to reproduce known expressions for the asymptotic tails of the momentum distributions and static structure factors of each system. Our conclusions are summarized in Section~\ref{sec:summary}, and several technical details are relegated to the Appendices.
\section{Factorization in the two-body system}
\label{sec:twobody}
In Ref.~\cite{Anderson:2010aq}, Anderson {\it et al.} applied renormalization group methods to the two-body problem to show that high-momentum components of low-energy wave functions factorize into the product of a state-independent function of momentum, and a state-dependent number that is sensitive only to low-momentum physics. Using this wave function factorization, it is straightforward to show that expectation values of operators that probe high-momentum modes similarly factorize into a state-independent piece that encodes the high-momentum physics and depends on the particular operator, and a state-dependent number that depends only on the low-momentum structure of the state and is identical for all high-momentum operators~\cite{Anderson:2010aq}. 
As we will show in Section~\ref{sec:Abodyfac}, the factorization formulas of Ref.~\cite{Anderson:2010aq} generalize to arbitrary low-energy $A$-body systems, allowing us to derive scaling relations for the high-momentum tails of momentum distributions and static structure factors. Since the simple factorization in the $A=2$ system is the starting point to derive analogous relations for general low-energy $A$-body states, we begin by reviewing the salient points from Ref.~\cite{Anderson:2010aq}.  

\subsection{Wave function factorization}

Renormalization group transformations simplify nuclear few- and many-body calculations by decoupling low- and high-momentum degrees of freedom leaving low-energy observables unchanged~\cite{Bogner:2003wn,Bogner:2009bt,Furnstahl:2012fn}. In Ref.~\cite{Anderson:2010aq}, the analysis was done in the context of similarity renormalization group (SRG) transformations, where the resolution scale $\Lambda$ provides a measure of how band-diagonal the evolved interaction is in momentum space\footnote{Using the SRG transformation of Ref.~\cite{Anderson:2010aq}, the evolved potential goes as $\frac{V^{\Lambda}(k',k)}{V^{\infty}(k',k)}\sim \exp\left(-\frac{(k'^2-k^2)^2}{\Lambda^4}\right)$.}.  However, for our present analysis we do not have to be very specific about the details of the particular RG implementation. All we require is that momentum modes above and below $\Lambda$ are effectively decoupled by the given transformation. In the center-of-mass frame of the two-body system, this implies that the low energy states ($|E_n| \lesssim \Lambda^2$) are localized in the low-momentum subspace\footnote{For SRG transformations, the high-momentum components of the evolved wave functions are exponentially suppressed as 
$\exp( -q^4/\Lambda^4)$. The decoupling is exact for RG transformations employing a sharp cutoff $\Lambda$. } 
\begin{eqnarray}
\projP_{\Lambda}|\psi^{\Lambda}_n\rangle &\approx &|\psi^{\Lambda}_n\rangle \qquad 
\projQ_{\Lambda}|\psi^{\Lambda}_n\rangle \approx  0\,,
\end{eqnarray}
where the projection operators $\projP_{\Lambda}$ and $\projQ_{\Lambda}$ are defined as 
\beqn
  \projP_{\Lambda} = \int^{\Lambda}_{0}\! \frac{d^3p}{(2\pi)^3} \ \vert \p{p} 
  \rangle\langle \p{p} \vert \quad {\rm and \quad}\projQ_{\Lambda} = \int^{\infty}_{\Lambda}\! \frac{d^3 q}{(2\pi)^3} \
    \vert  \p{q} \rangle\langle  \p{q}\vert\quad\,.
\eeqn

Starting from the unevolved Schr\"odinger equation written in block-matrix form
\beqn
  \begin{pmatrix}
    \projP_{\Lambda}\Hinf\projP_{\Lambda} & 
    \projP_{\Lambda}\Hinf\projQ_{\Lambda} \\
    \projQ_{\Lambda}\Hinf\projP_{\Lambda} & 
    \projQ_{\Lambda}\Hinf\projQ_{\Lambda} \\
  \end{pmatrix}
  \begin{pmatrix}
    \projP_{\Lambda}\psiinfalpha\ \\
    \projQ_{\Lambda}\psiinfalpha \\
  \end{pmatrix}
    = E\\ _{\alpha}\begin{pmatrix}\projP_{\Lambda}\psiinfalpha \\
        \projQ_{\Lambda}\psiinfalpha \\
  \end{pmatrix}
  \;,
\eeqn
we can solve for the high-momentum projection of any eigenstate as
\bea
  \projQ_{\Lambda}\left\vert \psiinfalpha\right>
  &=& (E_{\alpha}-\projQ_{\Lambda}\Hinf\projQ_{\Lambda})^{-1}
  \projQ_{\Lambda}\Hinf\projP_{\Lambda}
  \projP_{\Lambda}\left\vert \psiinfalpha \right\rangle
  \nonumber \\
    &=& (E_{\alpha}
     -\projQ_{\Lambda}\Hinf\projQ_{\Lambda})^{-1}\projQ_{\Lambda}\Vinf
     \projP_{\Lambda}\left\vert \psiinfalpha \right\rangle  
  \;,
\label{eq:Qpsi}
\eea
where we have used
\((\projP_{\Lambda})^{2}=\projP_{\Lambda}\), \(\Hinf=T+\Vinf\),
and \(\projQ_{\Lambda}T\projP_{\Lambda}=0\). For low-energy states \( \psiinfalpha\) such that
 \(|E_{\alpha}|\ll  {\rm Min}[|E_{\projQ H\projQ }|]\sim \Lambda^2\) (where \(E_{\projQ H\projQ }\) are the eigenvalues of \(\projQ_{\Lambda}\Hinf\projQ_{\Lambda}\)), we can neglect the \(E_{\alpha}\)
dependence in Eq.~\ref{eq:Qpsi}
 \beqn
 \psiinfalpha(\p{q})\,\approx\, -\int^{\infty}_{\Lambda}\!d\wt q'\,\int^{\Lambda}_{0}\!d\wt p\, \left\langle \p{q}\right\vert
 \frac{1}{\projQ_{\Lambda}\Hinf\projQ_{\Lambda}}\left| \p{q}' \right> \Vinf(\p{q'},\p{p})\,\psiinfalpha(\p{p})\,,
 \eeqn
where we've introduced the abbreviation $d\wt q \equiv \frac{d^3q}{(2\pi)^3}$.
Assuming for simplicity that $\psiinfalpha$ is an S-wave state and that the potential \(\Vinf(\p{q}',\p{p})\) is slowly varying with respect to $\p{p}$ compared to  \( \psiinfalpha(\p{p})\) in the region \(p<\Lambda\) and 
 \(q'\gg\Lambda\), we can factorize the low- and high-momentum physics by expanding%
\begin{eqnarray}\nonumber
   \int^{\Lambda}_{0} \! d\wt p \,
     \Vinf(\p{q}',\p{p})\psiinfalpha (\p{p})
      &\approx& 
      \Vinf(\p{q}',\p{p}')\vert_{\p{p}'=0}\times 
        \int^{\Lambda}_{0}\! d\wt p \,\psiinfalpha (\p{p})
      \\  && 
      \null + \left.\frac{1}{2}\frac{d^{2}}{dp^{'2}}\Vinf(\p{q}',\p{p}')
      \right\vert_{\p{p}'=0}\times \int^{\Lambda}_{0}\! d\wt p\,p^{2}\,\psiinfalpha (\p{p})\,+\cdots\,,
\label{eq:Taylor}
\end{eqnarray}
which gives 
\begin{eqnarray}
  \psiinfalpha(\p{q}) 
   &\approx&\,\gamma(\p{q};\Lambda)\, \int^{\Lambda}_{0}\! d\wt p\,\psiinfalpha(\p{p})\,\,+
    \,\,\eta(\p{q};\Lambda)\, \int^{\Lambda}_{0}\! d\wt p\,p^2\,\psiinfalpha(\p{p})\,+\, \ldots\,,
\end{eqnarray}
where the state-independent functions that carry the $\p{q}$-dependence are defined as
\beqn
\gamma(\p{q};\Lambda)\equiv-\int^{\infty}_{\Lambda}d\wt{q}'\,
 \left\langle \p{q}\right\vert
 \frac{1}{\projQ_{\Lambda}\Hinf\projQ_{\Lambda}}\left| \p{q}' \right> \Vinf(\p{q}',\p{0})
 \;,
 \label{eq:gammalambda}
\eeqn
\beqn
\eta(\p{q};\Lambda)\equiv-\frac{1}{2}\,\int^{\infty}_{\Lambda}d\wt{q}'\,
 \left\langle \p{q}\right\vert
 \frac{1}{\projQ_{\Lambda}\Hinf\projQ_{\Lambda}}\left| \p{q}' \right> \left.\frac{d^{2}}{dp^{'2}}\Vinf(\p{q}',\p{p}')
      \right\vert_{\p{p}'=0} \;.
 \label{eq:etalambda}
\eeqn

It is known empirically~\cite{Bogner:2009bt,Lepage:1997cs} 
that the low-momentum projections of the low-energy eigenstates of the bare and evolved Hamiltonians are related by a wave function renormalization factor \(\projP_{\Lambda}\left\vert
\psiinfalpha \right\rangle\approx Z_{\Lambda}\left\vert
\psilamalpha \right\rangle\), which reflects the fact that RG evolution does not modify long-distance physics.  Using that
\begin{eqnarray}
\int^{\Lambda}_{0}\!d\wt p\, \psiinfalpha(\p{p}) &\approx & Z_{\Lambda}\int^{\Lambda}_{0}\!d\wt p\, \psilamalpha(\p{p}) = \left.Z_{\Lambda}\,\psilamalpha(\p{r})\right\vert_{\p{r}=0}\,\equiv \,Z_{\Lambda}\,\psilamalpha(0) \\
\int^{\Lambda}_{0}\!d\wt p\, p^2\,\psiinfalpha(\p{p}) &\approx & Z_{\Lambda}\int^{\Lambda}_{0}\!d\wt p\, p^2\,\psilamalpha(\p{p}) = -\left.Z_{\Lambda}\,\nabla^2\psilamalpha(\p{r})\right\vert_{\p{r}=0}\,\equiv\,-Z_{\Lambda}\,\nabla^2\psilamalpha(0)\, ,
\end{eqnarray}
we obtain the momentum space version of Lepage's non-relativistic operator product expansion~\cite{Lepage:1997cs} relating the short-distance structure of the unevolved or ``bare'' wave functions to those of the low-energy effective theory
\beqn
 \psiinfalpha(\p{q}) \approx \,\gamma(\p{q};\Lambda)
  Z_{\Lambda}\psilamalpha (
  0)-\eta(\p{q};\Lambda)
  Z_{\Lambda}\,\nabla^2\psilamalpha (0)+\cdots\,.
\label{eq:OPE}
\eeqn
If we keep only the leading term in the expansion
\beqn
 \psiinfalpha(\p{q}) \approx \,\gamma(\p{q};\Lambda)
  Z_{\Lambda}\psilamalpha (
  0)\,= \gamma(\p{q};\Lambda)
  Z_{\Lambda}\int^{\Lambda}_0\! d\wt p\,\psilamalpha(\p{p})\,,
  \label{eq:OPE_LO}
\eeqn
we see that the high-momentum components of the low-energy
eigenstates are factorized into a state-independent function
\(\gamma(\p{q};\Lambda)\), which summarizes the short-distance behavior
of the wave function, and a state-dependent coefficient that probes the low-momentum structure of the state.  

\subsection{Effective operators and factorization} 
\label{sub:EffOp2b}
Given the wave function factorization in Eq.~\ref{eq:OPE_LO}, we can now derive analogous factorization formulas for expectation values of general operators in the $A=2$ system.  
Consider the expectation value of an operator $\Oop$ in a low-energy eigenstate of the unevolved Hamiltonian
\begin{eqnarray}
\langle \psiinfalpha|\Oop|\psiinfalpha\rangle &=&  \int_0^\Lambda\!\!d\wt p \int_0^\Lambda\!\!d\wt p' \,\psiinfalphastar(\p{p})O(\p{p},\p{p}')\psiinfalpha(\p{p}')\,+\,
\int_0^\Lambda\!\!d\wt p \int_\Lambda^\infty\!\!d\wt q\, \psiinfalphastar(\p{p})O(\p{p},\p{q})\psiinfalpha(\p{q}) \nonumber \\
&+& \int_\Lambda^\infty\!\!d\wt q \int_0^\Lambda\!\!d\wt p\, \psiinfalphastar(\p{q})O(\p{q},\p{p})\psiinfalpha(\p{p})\,+\,
\int_\Lambda^\infty\!\!d\wt q \int_\Lambda^\infty\!\!d\wt q'\, \psiinfalphastar(\p{q})O(\p{q},\p{q}')\psiinfalpha(\p{q}')\,, \nonumber\\
\end{eqnarray}
where we have explicitly separated the low- and high-momentum integrals in forming the matrix element.
Next, we insert Eq.~\ref{eq:OPE_LO} and $\psiinfalpha(\p{p})\approx Z_{\Lambda}\psilamalpha(\p{p})$ for the high- and low-momentum components of $\psiinfalpha$, respectively. Since the matrix elements $O(\p{p},\p{q})$ and $O(\p{q},\p{p})$ involve well-separated momenta, we perform a Taylor expansion about $\p{p}=0$ and keep only the leading term giving
\begin{eqnarray}
\langle \psiinfalpha|\Oop|\psiinfalpha\rangle &\approx&  Z^2_{\Lambda}\,\int^{\Lambda}_{0}\!\!d\wt p \int^{\Lambda}_{0}\!\!d\wt p' \,\psilamalphastar(\p{p})O(\p{p},\p{p}')\psilamalpha(\p{p}')\nonumber\\
&+&\,2Z^2_{\Lambda}|\psilamalpha(0)|^2 \int^{\infty}_{\Lambda}\!\!d\wt q\, O(0,\p{q})\gamma(\p{q};\Lambda) \nonumber \\
&+& Z^2_{\Lambda}|\psilamalpha(0)|^2\,\int^{\infty}_{\Lambda}\!\!d\wt q \int^{\infty}_{\Lambda}\!\!d\wt q'\, \gamma^*(\p{q};\Lambda)O(\p{q},\p{q}')\gamma(\p{q}';\Lambda)\,. \nonumber\\
\end{eqnarray}
Since the evolved wave functions $\psilamalpha(\p{k})$ have vanishing or exponentially suppressed support for $\p{k}>\Lambda$, we can re-write this as
\begin{equation}
\label{eq:Oeff1}
\langle \psiinfalpha|\Oop|\psiinfalpha\rangle \,\approx\, Z^2_{\Lambda}\langle \psilamalpha|\Oop|\psilamalpha\rangle + g^{(0)}(\Lambda)\,\langle  \psilamalpha|\delta^{(3)}(\p{r})|\psilamalpha\rangle\,,
\end{equation}
where the coupling $g^{(0)}(\Lambda)$ is defined as
\begin{eqnarray}
g^{(0)}(\Lambda)&\equiv& 2Z^2_{\Lambda}\,\int^{\infty}_{\Lambda}\!\!d\wt q\, O(0,\p{q})\gamma(\p{q};\Lambda) \,\nonumber\\&&\qquad+\,Z^2_{\Lambda}\,\int^{\infty}_{\Lambda}\!\!d\wt q \int^{\infty}_{\Lambda}\!\!d\wt q'\, \gamma^*(\p{q};\Lambda)O(\p{q},\p{q}')\gamma(\p{q}';\Lambda)\,.
\label{eq:g0}
\end{eqnarray}
Recalling that the consistently evolved effective operator is defined by
\beqn
\langle \psiinfalpha|\Oop|\psiinfalpha\rangle \,\equiv \, \langle \psilamalpha|\Oop_{\Lambda}|\psilamalpha\rangle\,,
\eeqn
we see from Eq.~\ref{eq:Oeff1} that 
\beqn
\Oop_{\Lambda}\, \approx\, Z^2_{\Lambda}\,\Oop \,+\, g^{(0)}(\Lambda)\,\delta^{(3)}(\p{r})\,+\,\ldots\,,
\eeqn
where the ``$\ldots$'' contains higher derivatives of delta functions that arise from the gradient terms in Eq.~\ref{eq:OPE} as well as higher-order terms in the expansion of $O(\p{q},\p{p})$ about $\p{p}=0$. In this way, we see that the RG-evolved operators take on a universal form; the effects of the integrated-out high-momentum modes are absorbed in a rescaling of the unevolved operator at the initial resolution scale, plus a series of local, state-independent corrections that take the form of a derivative expansion with $\Lambda$-dependent couplings~\cite{Lepage:1997cs,Felline:2003mi}. As stressed by Lepage~\cite{Lepage:1997cs}, the universal form of these local corrections is analogous to the multipole expansion in classical electromagnetism; just as multipole moments may be calculated from an underlying theory (e.g., the true charge and current densities) or extracted from a finite number of experimental data, the same holds true for the couplings $Z_{\Lambda}, g^{(0)}(\Lambda)$, etc. 

Let us now consider the implications of Eq.~\ref{eq:Oeff1} for operators that predominantly probe high-momentum components of low-energy states.  Since such operators have negligible strength at low-momentum $\projP_{\Lambda}\Oop\,\projP_{\Lambda}\approx 0$, 
the first term in Eq.~\ref{eq:Oeff1} vanishes, leaving
\beqn
\langle \psiinfalpha|\Oop|\psiinfalpha\rangle \,\approx\, g^{(0)}(\Lambda)\,\langle  \psilamalpha|\delta^{(3)}(\p{r})|\psilamalpha\rangle\,.
\label{eq:factorizedOp}
\eeqn
Therefore, the expectation value of {\it any} operator that probes the high-momentum structure of low-energy states factorizes into a state-independent piece, $g^{(0)}(\Lambda)$, that depends on the particular high-momentum operator via Eq.~\ref{eq:g0}, times a state-dependent number, $\langle  \psilamalpha|\delta^{(3)}(\p{r})|\psilamalpha\rangle$, that is the same for any high-momentum $\Oop$, and is only sensitive to the low-momentum structure of the state since $\projP_{\Lambda}|\psiinfalpha\rangle \approx Z_{\Lambda}|\psilamalpha\rangle$.
 
The momentum distribution $\hat{n}_{\p{q}} = a^{\dagger}_{\p{q}}a_{\p{q}}$ for $\p{q}\gg \Lambda$ is a prototypical example of an operator that is sensitive to the high-momentum structure of wave functions. Since $\hat{n}_{\p{q}} = |\p{q}\rangle\langle \p{q}|$ for the $A=2$ system in the center-of-mass frame, Eq.~\ref{eq:factorizedOp} becomes
\beqn
\label{eq:nqdeut}
\langle\psiinfalpha|\hat{n}_{\p{q}}|\psiinfalpha\rangle\, \approx\, \gamma^2(\p{q};\Lambda)\,Z^2_{\Lambda}\, |\psilamalpha(0)|^2\,.
\eeqn
We see that momentum distributions in {\it all} low-energy states ($|E_{\alpha}|\lesssim \Lambda^2$) in the $A=2$ system share the same $\p{q}$-dependence for $\p{q}\gtrsim\Lambda$. In fact, we will find in the following section that the factorization formula, Eq.~\ref{eq:nqdeut}, generalizes to arbitrary $A$-body systems.

\section{Factorization in the $A$-body system }
\label{sec:Abodyfac}
\subsection{Evolved creation and annihilation operators}
\label{sub:2ndquant}
In order to proceed beyond the $A=2$ system, it is convenient to recast the results from the previous section in a second-quantized language. As a first step, we examine how the Fock space creation and annihilation operators evolve under RG transformations. Suppressing non-essential spin and isospin indices, the transformed operators can be expanded on the original operator basis as 
\begin{equation}
\label{eq:adagger_lambda}
a^{(\Lambda)\dagger}_\p{q} = a^\dagger_\p{q}+\sum\limits_{\p{k}_1,\p{k}_2}C^{\Lambda}_\p{q}(\p{k}_1,\p{k}_2)a^\dagger_{\p{k}_1}a^\dagger_{\p{k}_2}a_{\p{k}_1+\p{k}_2-\p{q}}\,+\,\ldots\,\equiv\,a^\dagger_\p{q} + \delta a^{(\Lambda)\dagger}_\p{q}\, 
\end{equation}
where the ``$\ldots$'' contains higher-rank terms that are generated ($a^{\dagger}a^{\dagger}a^{\dagger} a a$, etc.) when the RG evolution is carried out beyond the two-body level, i.e., when induced 3- and higher-body interactions in $H^{\Lambda}$ are not truncated during the flow. Note that the form of the coupling function $C^{\Lambda}_{\p{q}}(\p{k}_1,\p{k}_2)$ can be constrained further by boosting both sides of Eq.~\ref{eq:adagger_lambda} and using Galilean invariance to write
\begin{eqnarray}
a^{(\Lambda)\dagger}_\p{q-P} &=& a^\dagger_\p{q-P}+\sum\limits_{\p{k}_1,\p{k}_2}C^{\Lambda}_\p{q}(\p{k}_1,\p{k}_2)a^\dagger_{\p{k}_1-\p{P}}a^\dagger_{\p{k}_2-\p{P}}a_{\p{k}_1+\p{k}_2-\p{q}-\p{P}}\nonumber \\
&=&a^\dagger_\p{q-P}+\sum\limits_{\p{k}_1,\p{k}_2}C^{\Lambda}_\p{q}(\p{k}_1+\p{P},\p{k}_2+\p{P})a^\dagger_{\p{k}_1}a^\dagger_{\p{k}_2}a_{\p{k}_1+\p{k}_2-\p{q}+\p{P}}\, \nonumber\\
&=&a^\dagger_\p{q-P}+\sum\limits_{\p{k}_1,\p{k}_2}C^{\Lambda}_\p{q-P}(\p{k}_1,\p{k}_2)a^\dagger_{\p{k}_1}a^\dagger_{\p{k}_2}a_{\p{k}_1+\p{k}_2-\p{q}+\p{P}}\,,
\end{eqnarray}
which implies
\beqn
\label{eq:Galilean}
C^{\Lambda}_\p{q}(\p{k}_1+\p{P},\p{k}_2+\p{P}) \,=\, C^{\Lambda}_\p{q-P}(\p{k}_1,\p{k}_2)\,.
\eeqn

In the following, we restrict our attention to the leading non-trivial term in Eq.~\ref{eq:adagger_lambda}. This corresponds to neglecting induced three- and higher-body interactions in $H^{\Lambda}$ since the coefficient function $C^{\Lambda}_{\p{q}}(\p{k}_1,\p{k}_2)$ is uniquely determined from the RG evolution in the two-body system. This can be seen by considering the following matrix element between the zero-particle vacuum (which does not evolve under the RG) and a two-body eigenstate in the bare and evolved theories
\begin{eqnarray}
\langle \psiinfalpha|a^{\dagger}_{\frac{\p{P}}{2}+\p{p}}a^{\dagger}_{\frac{\p{P}}{2}-\p{p}}|0\rangle\,&=&\, \langle\psilamalpha|a^{(\Lambda)\dagger}_{\frac{\p{P}}{2}+\p{p}}a^{(\Lambda)\dagger}_{\frac{\p{P}}{2}+\p{p}}|0\rangle\,\nonumber\\
&=& \langle\psilamalpha|a^{\dagger}_{\frac{\p{P}}{2}+\p{p}}a^{\dagger}_{\frac{\p{P}}{2}-\p{p}}|0\rangle  + \langle\psilamalpha|\delta a^{\dagger}_{\frac{\p{P}}{2}+\p{p}}a^{\dagger}_{\frac{\p{P}}{2}-\p{p}}|0\rangle \nonumber \\
&=&  \langle\psilamalpha|a^{\dagger}_{\frac{\p{P}}{2}+\p{p}}a^{\dagger}_{\frac{\p{P}}{2}-\p{p}}|0\rangle \,+\,\sum\limits_{\p{k}}C^{\Lambda}_{\p{P}/2+\p{p}}(\p{P}/2+\p{k},\p{P}/2-\p{k})\,\langle\psilamalpha |a^\dagger_{\frac{\p{P}}{2}+\p{k}}a^\dagger_{\frac{\p{P}}{2}-\p{k}}|0\rangle\nonumber\\
&=&  \langle\psilamalpha|a^{\dagger}_{\frac{\p{P}}{2}+\p{p}}a^{\dagger}_{\frac{\p{P}}{2}-\p{p}}|0\rangle \,+\,\sum\limits_{\p{k}}C^{\Lambda}_{\p{p}}(\p{k},-\p{k})\,\langle\psilamalpha |a^\dagger_{\frac{\p{P}}{2}+\p{k}}a^\dagger_{\frac{\p{P}}{2}-\p{k}}|0\rangle\,,
\end{eqnarray}
where $\delta a^{\dagger}\ket{0}=0$ was used in the second line and Eq.~\ref{eq:Galilean} was used in the last step. Since the dependence on the COM momentum $\p{P}$ cancels on both sides, we are left with
\beqn
\psiinfalphastar(\p{p}) = \psilamalphastar(\p{p}) + \sum\limits_{\p{k}} C^{\Lambda}_{\p{p}}(\p{k},-\p{k})\,\psilamalphastar(\p{k})\,,
\eeqn
which can be inverted using the completeness of the $\{\psilamalpha\}$ to give\footnote{It is helpful to think of the SRG or the Lee-Suzuki-Okubo similarity transformation method, where the decoupling of low- and high-momentum modes is accomplished by a unitary transformation. By unitarity, one has $\sum\limits_{\alpha}\ket{\psiinfalpha}\bra{\psiinfalpha} = \mathds{1} = \sum\limits_{\alpha} \ket{\psilamalpha}\bra{\psilamalpha}$, where the sum over $\alpha$ is unrestricted.} 
\beqn
\label{eq:W}
C^{\Lambda}_{\p{p}}(\p{k},-\p{k}) = \sum\limits_{\alpha} \langle\p{k}|\psilamalpha\rangle\langle\psiinfalpha|\p{p}\rangle - \delta_{\p{k},\p{p}}\,.
\eeqn
 
One can use the results of Section~\ref{sec:twobody} to evaluate two important limiting cases of Eq.~\ref{eq:W} that will prove useful below. First, consider $C^{\Lambda}_{\p{p}}(\p{p}',-\p{p}')$ for $p,p' \lesssim \Lambda$. Using $\projP_{\Lambda}|\psiinfalpha\rangle \approx Z_{\Lambda} |\psilamalpha\rangle$ for low-energy states and $\projP_{\Lambda}|\psilamalpha\rangle \approx 0$ for $|E_{\alpha}|\gtrsim \Lambda^2$, Eq.~\ref{eq:W} becomes
\begin{eqnarray}
\label{eq:Cpp}
C^{\Lambda}_{\p{p}}(\p{p}',-\p{p}')\,&\approx & Z_{\Lambda}\sum\limits_{|E_\alpha|\lesssim\Lambda^2} \langle\p{p}'|\psilamalpha\rangle\langle\psilamalpha|\p{p}\rangle - \delta_{\p{p}',\p{p}}\nonumber\\
&\approx & \bigl(Z_{\Lambda}-1\bigr)\,\delta_{\p{p}',\p{p}}\,.
\end{eqnarray}
In the last step, we used that the low-energy evolved eigenstates span the low-momentum subspace due to decoupling
\beqn
\label{eq:approxP}
\projP_{\Lambda} = \sum\limits_{\p{p}\leq \Lambda}|\p{p}\rangle\langle\p{p}| \,\approx\, \sum\limits_{|E_\alpha|\lesssim\Lambda^2}|\psilamalpha\rangle\langle\psilamalpha|\,.
\eeqn
The other important limiting case is $C^{\Lambda}_{\p{q}}(\p{p},-\p{p})$ for $p \lesssim \Lambda$ and $q\gtrsim \Lambda$.  Inserting Eq.~\ref{eq:OPE_LO} into Eq.~\ref{eq:W} then gives
\begin{eqnarray}
\label{eq:Cqp}
C^{\Lambda}_{\p{q}}(\p{p},-\p{p}) &\approx& Z_{\Lambda}\gamma(q;\Lambda)\sum\limits_{|E_\alpha|\lesssim\Lambda^2} \langle\p{p}'|\psilamalpha\rangle\langle\psilamalpha|\p{r=0}\rangle\nonumber\\
&\approx & Z_{\Lambda}\gamma(q;\Lambda)\,,
\end{eqnarray}
where Eq.~\ref{eq:approxP} was used in the final step. 

\subsection{Factorization for momentum distributions}

In Eq.~\ref{eq:nqdeut}, we found that the expectation value of the momentum distribution in low-energy two-body states factorizes for large $q\gtrsim \Lambda$. 
Using the second-quantized formulation of Section~\ref{sub:2ndquant}, we will now show that a similar factorization occurs for general low-energy $A$-body states.  We begin by considering the expectation value of the consistently evolved momentum distribution operator for $q\gg \Lambda$ in an arbitrary low-energy $A$-body state
\begin{eqnarray}
\label{eq:momdist}
n_{\p{q}}&=& \langle\psiinfalphaA|a^{\dagger}_{\p{q}}a_{\p{q}}|\psiinfalphaA\rangle = \langle\psilamalphaA|[a^{\dagger}_{\p{q}}a_{\p{q}}]^{(\Lambda)}|\psilamalphaA\rangle \nonumber\\
&=&  \langle\psilamalphaA|\bigl\{a^{\dagger}_{\p{q}}a_{\p{q}}+\delta a^{\!(\Lambda)\dagger}_{\p{q}}a_{\p{q}} + a^{\dagger}_{\p{q}}\delta a^{\!(\Lambda)}_{\p{q}} + \delta a^{\!(\Lambda)\dagger}_{\p{q}}\delta a^{\!(\Lambda)}_{\p{q}}\bigr\}  |\psilamalphaA\rangle\,.
\end{eqnarray}

This is an exact equality provided that a) the evolved Hamiltonian $H^{\Lambda}$ includes all induced $3\,$-, $4\,$-,$\,\ldots\,A$-body interactions generated by the RG evolution, b) all higher-order terms for $\delta a^{(\Lambda)\dagger}$ and $\delta a^{(\Lambda)}$ in Eq.~\ref{eq:adagger_lambda} are included, and c) all possible $1\,$-, $2\,$-, $\ldots$, $A$-body operators generated by the terms in the curly brackets are kept\footnote{An $A$-body operator is defined as a normal-ordered string of $A$ $a^{\dagger}$'s and $A$ $a$'s.}. However, since we are only interested in the high-momentum tail of Eq.~\ref{eq:momdist}, and since one expects induced 3-body and higher operators contributing to $H^{\Lambda}$ and $[a^{\dagger}_{\p{q}}a_{\p{q}}]^{(\Lambda)}$ to be subleading so long as one doesn't evolve too low in $\Lambda$~\cite{Bogner:2009bt}, we will neglect them.

In what follows, we assume $\Lambda$ is of the same order as the physical momentum scales that characterize $\psilamalphaA$ (e.g., the Fermi momentum, $k_F$, for homogenous systems, $\sqrt{m\omega/\hbar}$ for harmonically-trapped systems, etc.).  Due to decoupling, the low-energy states $\psilamalphaA$ have vanishingly small support at high momentum. Therefore, any term in  $[a^{\dagger}_{\p{q}}a_{\p{q}}]^{(\Lambda)}$ that annihilates a high-momentum particle from $|\psilamalphaA\rangle$ or $\langle\psilamalphaA|$ will be suppressed. In this limit, we find that Eq.~\ref{eq:momdist} becomes
\begin{eqnarray}
n_\p{q}&\approx& \bra{\psilamalphaA}\delta a^{(\Lambda)\dagger}_\p{q} \delta a^{(\Lambda)}_\p{q} \ket{\psilamalphaA} \nonumber \\
&=& \sum\limits_{\p{k,k',K,K'}}C^{\Lambda}_{\p{q}}\left(\frac{\p{K}}{2}+\p{k},\frac{\p{K}}{2}-\p{k}\right)C^\Lambda_{\p{q}}\left(\frac{\p{K}'}{2}+\p{k'},\frac{\p{K}'}{2}-\p{k'}\right)\nonumber\\
&&\qquad\qquad\qquad\times\,  \bra{\psilamalphaA}a^\dagger_{\frac{\p{K}}{2}+\p{k}}a^\dagger_{\frac{\p{K}}{2}-\p{k}}a_{\p{K-q}}a^\dagger_{\p{K'-q}}a_{\frac{\p{K'}}{2}+\p{k'}}a_{\frac{\p{K'}}{2}-\p{k'}}\ket{\psilamalphaA}\,\nonumber\\[.1in]
&=&\sum\limits_{\p{k,k',K}}C^{\Lambda}_{\p{q}}\left(\p{\frac{\p{K}}{2}+k}\p{\frac{\p{K}}{2}-k}\right)C^{\Lambda}_{\p{q}}\left(\p{\frac{\p{K}}{2}+k'},\p{\frac{\p{K}}{2}-k'}\right)\bra{\psilamalphaA}a^\dagger_{\frac{\p{K}}{2}+\p{k}}a^\dagger_{\frac{\p{K}}{2}-\p{k}}a_{\frac{\p{K}}{2}+\p{k'}}a_{\frac{\p{K}}{2}-\p{k'}}\ket{\psilamalphaA}\nonumber\\
&=& \sum\limits_{\p{k,k',K}}C^{\Lambda}_{\p{q}-\frac{\p{K}}{2}}\left(\p{k},\p{-k}\right)C^{\Lambda}_{\p{q}-\frac{\p{K}}{2}}\left(\p{k'},\p{-k'}\right)\bra{\psilamalphaA}a^\dagger_{\frac{\p{K}}{2}+\p{k}}a^\dagger_{\frac{\p{K}}{2}-\p{k}}a_{\frac{\p{K}}{2}+\p{k'}}a_{\frac{\p{K}}{2}-\p{k'}}\ket{\psilamalphaA}\,,
\end{eqnarray}
where we have anti-commuted $a_{\p{K}-\p{q}}$ to the right and dropped the normal-ordered three-body term in going from the second to third line. The low-momentum nature of $\psilamalphaA$ implies that dominant terms in the sum are for $|\p{K}/2 \pm\p{k}|$ and $|\p{K}/2 \pm\p{k}'| \lesssim \Lambda$. Consequently, we have a mismatch of scales $|\p{q}-\p{K}/2| \gg |\p{k}|,|\p{k}'|$, which together with Eq.~\ref{eq:Cqp} gives
\begin{eqnarray}
\label{eq:momdistfacA}
n_{\p{q}}&\approx &Z_{\Lambda}^2 \sum\limits_{\p{k,k',K}} \gamma^2(\p{q}-\p{K}/2;\Lambda) \, \bra{\psilamalphaA}a^\dagger_{\frac{\p{K}}{2}+\p{k}}a^\dagger_{\frac{\p{K}}{2}-\p{k}}a_{\frac{\p{K}}{2}+\p{k'}}a_{\frac{\p{K}}{2}-\p{k'}}\ket{\psilamalphaA}\nonumber\\
&\approx &Z_{\Lambda}^2 \gamma^2(\p{q};\Lambda) \, \sum\limits_{\p{k,k',K}}\bra{\psilamalphaA}a^\dagger_{\frac{\p{K}}{2}+\p{k}}a^\dagger_{\frac{\p{K}}{2}-\p{k}}a_{\frac{\p{K}}{2}+\p{k'}}a_{\frac{\p{K}}{2}-\p{k'}}\ket{\psilamalphaA}\,,
\end{eqnarray}
where we've used $q\gg K/2$ in the last step\footnote{For systems where $\gamma(\p{q};\Lambda)$ exhibits a power-law decay, the corrections for non-zero $K$ do not modify the power-law tail of $n_{\p{q}}$.}. In this way, we see the large-$q$ tails of  momentum distributions for arbitrary low-energy $A$-body states share the same universal $q$-dependence. In nuclear physics, Eq.~\ref{eq:momdistfacA} provides an alternative to the usual explanations based on short-range correlations~\cite{Frankfurt:2008zv,Frankfurt:2009vv} as to why calculated momentum distributions in various nuclei and nuclear matter scale with each other at large $q$. 
In Section~\ref{sec:examples}, we will use Eq.~\ref{eq:momdistfacA} and the analogous expression, Eq.~\ref{eq:staticEff2}, to reproduce known asymptotic expressions for the momentum distributions and static structure factors for two well-studied many-body systems, the unitary Fermi gas and the electron gas. 

\subsection{Factorization for static structure factors}
The static structure factor is an important quantity that contains information about density-density correlations in a many-body system. For a many-body system of fermions with two spin states, the correlations between the densities of the two spin states are particularly important.
The corresponding static structure factor $S_{\uparrow\!\downarrow}(\p{q})$ for a homogeneous system
is the Fourier transform in the relative coordinate $\p{r}_1-\p{r}_2$ of the density correlator
$\langle\psiinfalphaA| \rho_{\uparrow}(\p{r}_1)\rho_{\downarrow}(\p{r}_2)|\psiinfalphaA\rangle$. Using similar arguments as for the momentum distribution, we now show that at large momentum $S_{\uparrow\!\downarrow}(\p{q})$ factorizes into a universal function of $\p{q}$ times a matrix element of a delta function in the evolved low-momentum wave functions. 

Starting from the definition of $S_{\uparrow\!\downarrow}(\p{q})$ in the unevolved theory
\begin{eqnarray}
\label{StrucBas}
S_{\uparrow\!\downarrow}(\p{q})&=&\bra{\psiinfalphaA}\rho_\uparrow^\dagger(\p{q})\rho_{\downarrow}(\p{q})\ket{\psiinfalphaA} = \sum\limits_{\p{p,p'}}\bra{\psiinfalphaA}a^\dagger_{\p{p},\uparrow}a_{\p{p+q},\uparrow}a^\dagger_{\p{p'+q},\downarrow}a_{\p{p'},\downarrow}\ket{\psiinfalphaA} \nonumber \\
&=& \sum\limits_{\p{p,p'}}\bra{\psiinfalphaA}a^{\dagger}_{\p{p'+q},\downarrow}a^\dagger_{\p{p},\uparrow}
a_{\p{p+q},\uparrow}a_{\p{p'},\downarrow}\ket{\psiinfalphaA}\nonumber \\
&\equiv& \bra{\psiinfalphaA}\widehat{S}_{\uparrow\!\downarrow}(\p{q})\ket{\psiinfalphaA}\,,
\end{eqnarray}
we consider the expectation value of the consistently evolved operator $\widehat{S}_{\uparrow\!\downarrow}(\p{q};\Lambda)$ in the evolved wave functions. Using Eq.~\ref{eq:adagger_lambda} for the evolved creation/annihilation operators, we have
\begin{eqnarray}
\label{eq:staticEff}
\bra{\psiinfalphaA}\widehat{S}_{\uparrow\!\downarrow}(\p{q})\ket{\psiinfalphaA} &=& \bra{\psilamalphaA}\widehat{S}_{\uparrow\!\downarrow}(\p{q};\Lambda)\ket{\psilamalphaA} \equiv \bra{\psilamalphaA}\bigl(\widehat{S}_{\uparrow\!\downarrow}(\p{q})+\delta\widehat{S}_{\uparrow\!\downarrow}^{\Lambda}(\p{q})\bigr)\ket{\psilamalphaA}\,.
\end{eqnarray}
This is an exact relation only if all induced many-body operators (up to rank-$A$ for the $A$-body system) are kept in $H^{\Lambda}$ and $\delta\widehat{S}^{\Lambda}(\p{q})$. As with our analysis of the momentum distribution, we neglect these many-body contributions by a) restricting the expansion of $a^{(\Lambda)\dagger}$ and $a^{(\Lambda)}$ to the leading terms shown in Eq.~\ref{eq:adagger_lambda} and, b) truncating $\delta\widehat{S}^{\Lambda}(\p{q})$ to two-body operators
\begin{eqnarray}
\delta\widehat{S}_{\uparrow\!\downarrow}^{\Lambda}(\p{q}) &\approx& \sum\limits_{\p{K,k,k'}}C^{\Lambda}_{\p{q}+\p{k}'}(\p{k},-\p{k})\,a^\dagger_{\frac{\p{K}}{2}+\p{k},\uparrow}a^\dagger_{\frac{\p{K}}{2}-\p{k},\downarrow}a_{\frac{\p{K}}{2}+\p{k'},\downarrow}a_{\frac{\p{K}}{2}-\p{k'},\uparrow} \,+\, \rm{h.c.}\nonumber\\
&+&\,\sum\limits_{\p{P,K,k,k'}}C^{\Lambda}_{\p{P+q-\frac{K}{2}}}(\p{k},-\p{k})C^{\Lambda}_{\p{P-\frac{K}{2}}}(\p{k}',-\p{k}')\,a^\dagger_{\frac{\p{K}}{2}+\p{k},\uparrow}a^\dagger_{\frac{\p{K}}{2}-\p{k},\downarrow}a_{\frac{\p{K}}{2}+\p{k'},\downarrow}a_{\frac{\p{K}}{2}-\p{k'},\uparrow}\,\nonumber\\
&\equiv& \delta\widehat{S}_{1}^{\Lambda}(\p{q})\,+\, \delta\widehat{S}_{2}^{\Lambda}(\p{q})\,,
\end{eqnarray}
where $ \delta\widehat{S}_{1(2)}^{\Lambda}(\p{q})$ denotes the terms linear (quadratic) in the expansion coefficients $C^{\Lambda}$. 

With the approximate form of the evolved operator in hand, we can now evaluate Eq.~\ref{eq:staticEff} for $q \gg \Lambda$, where once again $\Lambda$ is assumed to be of the same order as the physical scales that characterize the system. The expectation value of the bare operator, $\hat{S}_{\uparrow\!\downarrow}(\p{q})$, in the evolved low-momentum wave functions is negligible since it involves the removal of a high-momentum particle. Therefore, we have
\begin{equation}
\begin{aligned}
\label{eq:staticEff1a}
\bra{\psiinfalphaA}\widehat{S}_{\uparrow\!\downarrow}(\p{q})\ket{\psiinfalphaA} &\approx\bra{\psilamalphaA}\bigl(\delta\widehat{S}_{1}^{\Lambda}(\p{q})+\delta\widehat{S}_{2}^{\Lambda}(\p{q})\bigr)\ket{\psilamalphaA} \\
&= 2\,\sum\limits_{\p{K,k,k'}}C^{\Lambda}_{\p{q}+\p{k}'}(\p{k},-\p{k})\, \bra{\psilamalphaA}a^\dagger_{\frac{\p{K}}{2}+\p{k},\uparrow}a^\dagger_{\frac{\p{K}}{2}-\p{k},\downarrow}a_{\frac{\p{K}}{2}+\p{k'},\downarrow}a_{\frac{\p{K}}{2}-\p{k'},\uparrow}\ket{\psilamalphaA}  \\
+ \sum\limits_{\p{P,K,k,k'}}&C^{\Lambda}_{\p{P+q-\frac{K}{2}}}(\p{k},-\p{k})C^{\Lambda}_{\p{P-\frac{K}{2}}}(\p{k}',-\p{k}')\,\bra{\psilamalphaA}a^\dagger_{\frac{\p{K}}{2}+\p{k},\uparrow}a^\dagger_{\frac{\p{K}}{2}-\p{k},\downarrow}a_{\frac{\p{K}}{2}+\p{k'},\downarrow}a_{\frac{\p{K}}{2}-\p{k'},\uparrow}\,\ket{\psilamalphaA}\,.
\end{aligned}
\end{equation}
Due to the low-momentum structure of the evolved wave functions, it is clear that the sums over momenta $\p{K},\p{k},\p{k}'$ are effectively cutoff at $\Lambda$, while the summation over $\p{P}$ is unrestricted in the second term of Eq.~\ref{eq:staticEff1a}. Performing a Taylor series expansion of the coefficient functions in powers of the small momenta $\p{K},\p{k},\p{k}'$ and keeping just the leading term gives
\begin{eqnarray}
\label{eq:staticEff1}
\bra{\psiinfalphaA}\widehat{S}_{\uparrow\!\downarrow}(\p{q})\ket{\psiinfalphaA} &\approx& 
\Bigl\{2\,C^{\Lambda}_{\p{q}}(0,0) 
+ \sum\limits_{\p{P}}C^{\Lambda}_{\p{P+q}}(0,0)C^{\Lambda}_{\p{P}}(0,0)\,\Bigr\}\nonumber\\
&&\qquad\qquad\qquad\times\quad
\sum\limits_{\p{K,k,k'}}\, \bra{\psilamalphaA}a^\dagger_{\frac{\p{K}}{2}+\p{k},\uparrow}a^\dagger_{\frac{\p{K}}{2}-\p{k},\downarrow}a_{\frac{\p{K}}{2}+\p{k'},\downarrow}a_{\frac{\p{K}}{2}-\p{k'},\uparrow}\ket{\psilamalphaA}\,.\nonumber\\
\end{eqnarray}
To proceed further, we consider the following three regions that arise in the sum over $\p{P}$: 
\begin{itemize}
\item Region I): $|\p{P}+\p{q}| \gtrsim \Lambda$ and $|\p{P}| \gtrsim \Lambda$
\item Region II): $|\p{P}+\p{q}|\lesssim \Lambda$ and $|\p{P}| \gtrsim \Lambda$ 
\item Region III): $|\p{P}+\p{q}| \gtrsim \Lambda$ and $|\p{P}| \lesssim \Lambda$. 
\end{itemize}

Regions II) and III) are trivial since the $C^{\Lambda}$ coefficients involving all soft momenta give a delta function, Eq.~\ref{eq:Cpp}, that allows the sums to be performed. Together with Eq.~\ref{eq:Cqp}, we have 
\begin{eqnarray}
&&\sum\limits_{\p{P},\rm{II}}C^{\Lambda}_{\p{P+q}}(0,0)C^{\Lambda}_{\p{P}}(0,0)\, \approx \sum\limits_{\p{P},\rm{II}}(Z_{\Lambda}-1)\delta_{\p{P},\p{q}}\,Z_{\Lambda}\,\gamma(\p{P};\Lambda)\,=\,Z_{\Lambda}(Z_{\Lambda}-1)\gamma(\p{q};\Lambda)\nonumber\\
&&\sum\limits_{\p{P},\rm{III}}C^{\Lambda}_{\p{P+q}}(0,0)C^{\Lambda}_{\p{P}}(0,0)\, \approx \sum\limits_{\p{P},\rm{III}}(Z_{\Lambda}-1)\delta_{\p{P},\p{0}}\,Z_{\Lambda}\,\gamma(\p{P+q};\Lambda)\,=\,Z_{\Lambda}(Z_{\Lambda}-1)\gamma(\p{q};\Lambda)\,,\nonumber\\
\end{eqnarray}
which gives 
\begin{eqnarray}
\label{eq:staticEff2}
\bra{\psiinfalphaA}\widehat{S}_{\uparrow\!\downarrow}(\p{q})\ket{\psiinfalphaA} &\approx& 
\Bigl\{2\,Z_{\Lambda}^2\gamma(\p{q};\Lambda)
+ \sum\limits_{\p{P},I}Z_{\Lambda}^2\,\gamma(\p{P+q};\Lambda)\gamma(\p{P};\Lambda)\,\Bigr\}\nonumber\\
&&\qquad\qquad\qquad\times\quad
\sum\limits_{\p{K,k,k'}}\, \bra{\psilamalphaA}a^\dagger_{\frac{\p{K}}{2}+\p{k},\uparrow}a^\dagger_{\frac{\p{K}}{2}-\p{k},\downarrow}a_{\frac{\p{K}}{2}+\p{k'},\downarrow}a_{\frac{\p{K}}{2}-\p{k'},\uparrow}\ket{\psilamalphaA}\,.\nonumber\\
\end{eqnarray} 
As with the momentum distribution, we see that the high-momentum tail of the static structure factor in a general low-energy $A$-body state factorizes into a universal function of $\p{q}$, multiplied by a state-dependent matrix element that is controlled entirely by low-momentum physics. 

\subsection{Factorization for general high-momentum operators}

While our explicit proofs of factorization have thus far been limited to the momentum distribution and the static structure factor, the phenomena is very general and can be qualitatively understood from Eq.~\ref{eq:Oeff}. Consider an operator at the initial high-resolution scale $\Lambda_0$ that probes high-momentum modes, $\hat{O}^{\Lambda_0}_{\p{q}}$, where the subscript $\p{q}$ indicates that the second-quantized expression involves creation (annihilation) operators that add (remove) a high-momentum particle. We assume $\p{q}$ is much larger than any physical scale that characterizes the low-energy state $\psi^{\Lambda_0}_{n}$, and we also assume $|\p{q}|\ll \Lambda_0$ so that the expectation value of $\hat{O}^{\Lambda_0}_{\p{q}}$ is non-vanishing. Now consider the consistently evolved $\hat{O}^{\Lambda}_{\p{q}}$, where $\Lambda \ll |\p{q}|\ll \Lambda_0$, and expand it as a polynomial in creation/annihilation operators defined at $\Lambda_0$. Schematically, we have
\begin{equation}
\hat{O}^{\Lambda}_{\p{q}} = \sum_{\alpha} g^{\alpha}_{\p{q}} \hat{A}_{\alpha}\,
\end{equation}
where $\hat{A}_{\alpha}$ denotes a normal-ordered string of creation/annihilation operators at $\Lambda_0$, $\alpha$ is a collective index for the different momentum modes being added/removed, and $g^{\alpha}_{\p{q}}$ is a c-number coefficient. Inserting this into Eq.~\ref{eq:Oeff}, we have
\begin{equation}
\label{eq:facgen}
\bra{\psi_n^{\Lambda_0}}\hat{O}^{\Lambda_0}_{\p{q}}\ket{\psi_n^{\Lambda_0}} = \sum_{\alpha}g^{\alpha}_{\p{q}}\,\bra{\psi_n^{\Lambda}}\hat{A}_{\alpha}\ket{\psi_n^{\Lambda}}\,.
\end{equation}
Due to the low-momentum nature of the evolved wave functions, we find that only the $\hat{A}_{\alpha}$  involving the addition/removal of low-momentum ($\lesssim \Lambda$) modes contributes to Eq.~\ref{eq:facgen}. Since all momentum modes contained in $\alpha$ obey $k_{\alpha}/q \ll 1$, the c-number coefficients can be Taylor-expanded in the soft-momenta. Note that this expansion should be well-defined since the $g^{\alpha}_{\p{q}}$ encode contributions from loop integrals that have both ultraviolet ($\Lambda_0$) and infrared ($\Lambda$) cutoffs in place, thus preventing any singular behavior from arising. In this way, we see that the universal $\p{q}$-dependence factorizes, and the remaining state-dependence is given by matrix elements of low-momentum operators.

\section{Examples}
\label{sec:examples}
As a check of our factorization formulas Eq.~\ref{eq:momdistfacA} and Eq.~\ref{eq:staticEff2}, we use them to reproduce known expressions for the high-momentum tails of  $n_{\p{q}}$ and $S_{\uparrow\!\downarrow}(\p{q})$ for two well-studied systems, the unitary Fermi gas (UFG) and the electron gas.  
\subsection{Unitary Fermi gas}
\subsubsection{Momentum distribution}
In the case of a unitary Fermi gas described by a contact interaction, the coefficient $\gamma(\p{q};\Lambda)$, and hence the large momentum tails of $n_{\p{q}}$ and $S(\p{q})$, can be calculated analytically. Consider the two-body Hamiltonian with a spin-independent contact interaction
\begin{equation}
\hat{H}_\infty =\hat{T} + \hat{V_\delta} =\frac{\hat{\p{p}}^2}{2m}+\frac{g(\Lambda_0)}{2m}\delta^{(3)}(\p{r})\,,\end{equation}
where $\Lambda_0$ is the ultraviolet cutoff on all momenta of the theory. Here, we assume that $\Lambda_0$ is much larger than any relevant low-energy scales in the problem such as the inverse scattering length or the Fermi momentum. The coupling constant $g(\Lambda_0)$ is determined by matching the scattering amplitude at threshold to the $S$-wave scattering length $a$ and is given by~\cite{Braaten:2008uh} 
\begin{equation}
\label{Coupling}
g(\Lambda_0) = \left[\frac{1}{4\pi a}-\frac{\Lambda_0}{2\pi^2}\right]^{-1}\,.
\end{equation}
To obtain an explicit expression for $\gamma(\p{q};\Lambda)$ in Eq.~\ref{eq:gammalambda}, the operator $(\projQ_{\Lambda} H_\infty \projQ_{\Lambda})^{-1}$ can be constructed with the aid of the operator identity
\begin{equation}
\frac{1}{A+B} = (1-A^{-1}B+A^{-1}BA^{-1}B-\ldots)A^{-1}\,,
\end{equation}
where $A\rightarrow \projQ_{\Lambda}T\projQ_{\Lambda}$ and $B\rightarrow \projQ_{\Lambda}V_\delta\projQ_{\Lambda}$ giving
\begin{equation}\begin{aligned}
\gamma(\p{q};\Lambda) &= -\frac{g(\Lambda_0)}{2m}\int_{\Lambda}^{\Lambda_0}\frac{\dif^3{q'}}{(2\pi)^3}\bra{\p{q}}(\projQ_{\Lambda}H_\infty \projQ_{\Lambda})^{-1}\ket{\p{q'}} \\
&=-\frac{g(\Lambda_0)}{q^2}\left(1-\frac{g(\Lambda_0)}{q^2}\int_{\Lambda}^{\Lambda_0}\frac{\dif{q'}}{2\pi^2}\frac{q'^2}{q'^2}+\ldots\right) = -\frac{g(\Lambda_0)}{q^2}\sum\limits_{n=0}^\infty\left(-\frac{g(\Lambda_0)\cdot(\Lambda_0-\Lambda)}{2\pi^2}\right)^n\\
&=\frac{-g(\Lambda_0)}{q^2}\frac{2\pi^2}{2\pi^2+g(\Lambda_0)\cdot(\Lambda_0-\Lambda)}\,.
\end{aligned}\end{equation}
This can be simplified further since Eq.~\ref{Coupling} implies
\begin{equation}
\frac{2\pi^2g(\Lambda_0)}{2\pi^2+g(\Lambda_0)\cdot(\Lambda_0-\Lambda)} = \left[\frac{\Lambda_0-\Lambda}{2\pi^2}+\frac{1}{g(\Lambda_0)}\right]^{-1}=\left[\frac{1}{4\pi a}-\frac{\Lambda}{2\pi^2}\right]^{-1} = g(\Lambda)\,,
\end{equation}
which gives
\begin{equation}
\label{eq:gammaUFG}
\gamma(\p{q};\Lambda) = -\frac{g(\Lambda)}{q^2}\,.
\end{equation}
Inserting into Eq.~\ref{eq:momdistfacA}, we find 
\begin{equation}
\label{eq:momdistfacUFG}
n_{\p{q}}\approx \frac{Z_{\Lambda}^2g^2(\Lambda)}{q^4}\sum\limits_{\p{k,k',K}}\bra{\psilamalphaA}a^\dagger_{\frac{\p{K}}{2}+\p{k}}a^\dagger_{\frac{\p{K}}{2}-\p{k}}a_{\frac{\p{K}}{2}+\p{k'}}a_{\frac{\p{K}}{2}-\p{k'}}\ket{\psilamalphaA}\,,
\end{equation}
where $\Lambda$ is of the same order of magnitude as the relevant low-energy scales of the system and $\Lambda \ll q \ll \Lambda_0$. 

In Ref.~\cite{Braaten:2008uh}, Braaten and Platter used the  operator product expansion to show that the tail of the UFG momentum distribution behaves like\footnote{Shina Tan provided the first derivation of Eq.~\ref{eq:momdistOPE} using generalized functions~\cite{2008AnPhy.323.2971T}. Many different derivations can be found in the literature, see Ref.~\cite{Braaten:2010if} for details.}
\begin{equation}
\label{eq:momdistOPE}
n_\p{q} = \frac{g^2(\Lambda_0)}{q^4}\sum\limits_{\p{k,k',K}}\bra{\psi_{\alpha,_A}^{\Lambda_0}}a^\dagger_{\frac{\p{K}}{2}+\p{k}}a^\dagger_{\frac{\p{K}}{2}-\p{k}}a_{\frac{\p{K}}{2}+\p{k'}}a_{\frac{\p{K}}{2}-\p{k'}}\ket{\psi_{\alpha,_A}^{\Lambda_0}} \equiv \frac{C(\Lambda_0)}{q^4}
\end{equation}
where $C(\Lambda_0)$ is often known as Tan's contact parameter. In the Appendix, we will show that, at the level of approximating the evolved creation/annihilation operators by the leading-order expression in Eq.~\ref{eq:adagger_lambda} and truncating induced three- and higher-body operators, the following relationship holds
\begin{eqnarray}
\label{eq:contact}
Z_{\Lambda}^2 C(\Lambda) 
= C(\Lambda_0)\,.
\end{eqnarray}
Therefore, Eqs.~\ref{eq:momdistfacUFG} and \ref{eq:momdistOPE} are equivalent at the level of approximations made thus far. Heuristically, we can understand this equivalence since we expect that $Z_{\Lambda}\rightarrow 1$ as $\Lambda\rightarrow\Lambda_0$.

\subsubsection{Static structure factor}
Turning next to the asymptotic expression for the static structure factor, Eq.~\ref{eq:staticEff2}, 
our task is to evaluate the following term 
\begin{equation}
\label{eq:UFGintegral1}
\sum\limits_{\p{P},I}Z_{\Lambda}^2\,\gamma(\p{P+q};\Lambda)\gamma(\p{P};\Lambda)\,\,\rightarrow\,\,Z_{\Lambda}^2g^2(\Lambda)\,\int \frac{d^3P}{(2\pi)^3}\frac{\theta(|\p{P}|-\Lambda)\theta(|\p{P}+\p{q}|-\Lambda)}{|\p{P}+\p{q}|^2|\p{P}|^2}\,,
\end{equation}
where we've taken the infinite volume limit to convert the sum to an integral and substituted Eq.~\ref{eq:gammaUFG} for $\gamma$.  Note that the integral in Eq.~\ref{eq:UFGintegral1} has an implicit ultra-violet cutoff $\Lambda_0\gg \Lambda$. To evaluate the integral,  we use that there are two regions for which the theta function $\theta(|\p{P}+\p{q}|-\Lambda)=1$ independent of $\p{P}\cdot\p{q}$, and one region where it depends on angle to write 
\begin{equation}
\label{eq:IUFG}
I = \int \frac{d^3P}{(2\pi)^3}\frac{\theta(|\p{P}|-\Lambda)\theta(|\p{P}+\p{q}|-\Lambda)}{|\p{P}+\p{q}|^2|\p{P}|^2}\, \equiv I_{\rm{high}}\,+\,I_{\rm{medium}}\,+\, I_{\rm{low}}\,.
\end{equation}

For $|\p{P}|>|\p{q}|$, we have $|\p{P}+\p{q}|\ge |\p{P}|-|\p{q}|$, which implies that for $|\p{P}|\ge \Lambda + |\p{q}|$, then $\theta(|\p{P}+\p{q}|-\Lambda)=1$ independent of $\p{P}\cdot\p{q}$. In this case, the limits of the angular integration are unrestricted
\begin{eqnarray}
\label{eq:IhighUFG}
I_{\rm{high}}&=& \frac{1}{4\pi^2}\int_{\Lambda+q}^{\Lambda_0}dP\,\int_{-1}^{1}dx\frac{1}{P^2+q^2 + 2Pqx}\,
\overset{\Lambda_0\rightarrow\infty}{=}\frac{1}{4\pi^2q}\left[\Li_2\left(\frac{q}{q+\lambda}\right)-\Li_2\left(-\frac{q}{q+\lambda}\right)\right]\,,\nonumber\\
\end{eqnarray}
where $\rm{Li}_2(x)$ is the polylogarithm function. Using that $\Lambda/q \ll 1$ and keeping just the leading term gives
\begin{equation}
\label{eq:IhighUFG1}
I_{\rm{high}} \approx \frac{1}{4\pi^2q}\,\biggl[\frac{\pi^2}{4}\,+\,\frac{\Lambda}{q}\log\left(\frac{\Lambda}{2q}\right)\,
-\,\frac{\Lambda}{q}\biggr]\,. 
\end{equation}
Similarly, for $|\p{P}|<|\p{q}|$ the limits of the angular integration are unrestricted if $ |\p{P}|< |\p{q}|-\Lambda$ giving 
\begin{eqnarray}
\label{eq:IlowUFG}
I_{\rm{low}}&=& \frac{1}{4\pi^2}\int_{\Lambda}^{q-\Lambda}dP\,\int_{-1}^{1}dx\frac{1}{P^2+q^2 + 2Pqx}\nonumber\\
&=&\frac{1}{4\pi^2q}\left[\Li_2\left(-\frac{\Lambda}{q}\right)-\Li_2\left(\frac{\Lambda}{q}\right) +\Li_2\left(1-\frac{\Lambda}{q}\right)-\Li_2\left(-1+\frac{\Lambda}{q}\right)\right]\,,\nonumber\\
&\approx&  \frac{1}{4\pi^2q}\,\biggl[\frac{\pi^2}{4}\,+\,\frac{\Lambda}{q}\log\left(\frac{\Lambda}{2q}\right)\,
-\,\frac{3\Lambda}{q}\biggr]\,,
\end{eqnarray}
where we've used $\Lambda/q \ll 1$ in the last step. Finally, we consider the intermediate region $q-\Lambda < P < q+\Lambda$ where $\theta(|\p{P}+\p{q}|-\Lambda)=1$ places restrictions on the the limits of the angular integration. In this case, the theta function requires $x > x_{\rm{min}}= \frac{\Lambda^2-P^2-q^2}{2Pq}$ 
\begin{eqnarray}
\label{eq:ImedUFG}
I_{\rm{medium}} &=&  \frac{1}{4\pi^2}\int_{q-\Lambda}^{q+\Lambda}dP\,\int_{x_{\rm min}}^{1}dx\frac{1}{P^2+q^2 + 2Pqx}\nonumber\\
&=&\frac{1}{4\pi^2q}\,\biggl[\log\left(\frac{q}{\Lambda}\right)\log\left(\frac{1+\Lambda/q}{1-\Lambda/q}\right)\,+\,\Li_2\left(-1-\frac{\Lambda}{q}\right)-\Li_2\left(-1+\frac{\Lambda}{q}\right)\biggr]\nonumber\\
&\approx&  \frac{1}{4\pi^2q}\,\biggl[\frac{2\Lambda}{q}\log\left(\frac{2q}{\Lambda}\right)\biggr]\,.
\end{eqnarray}
Inserting Eqs.~\ref{eq:IhighUFG1}-\ref{eq:ImedUFG} in Eq.~\ref{eq:IUFG} gives
\begin{equation}
I \approx \frac{1}{4\pi^2q}\left[\frac{\pi^2}{2}-\frac{4\Lambda}{q}\right]\,,
\end{equation}
which together with Eq.~\ref{eq:staticEff2} and Eq.~\ref{eq:gammaUFG} yields
\begin{eqnarray}
\label{eq:staticEff3}
S_{\uparrow\!\downarrow}(\p{q}) &\approx& \left(-\frac{2}{q^2 g(\Lambda)} \,+\,\frac{1}{8q}\,-\,\frac{\Lambda}{\pi^2q^2}\right)\,Z_{\Lambda}^2\,C(\Lambda)\,\nonumber\\
&=& \left(-\frac{2}{q^2 g(\Lambda)} \,+\,\frac{1}{8q}\,-\,\frac{\Lambda}{\pi^2q^2}\right)\,C(\Lambda_0) \nonumber\\
&=& \left(\frac{1}{8q} \,-\,\frac{1}{2\pi a q^2}\right)\,C(\Lambda_0)\,,
\end{eqnarray}
where we used Eq.~\ref{eq:contact} and the explicit form of the coupling $g(\Lambda)$, Eq.~\ref{Coupling}, in the second and third lines, respectively. As with the momentum distribution, Eq.~\ref{eq:staticEff3} agrees with the known result that has been previously derived by a number of different methods~\cite{Braaten:2010if}.

\subsection{Electron gas}

\subsubsection{Momentum distribution}
As our second check of Eq.~\ref{eq:momdistfacA} and Eq.~\ref{eq:staticEff2}, we derive the large-momentum limit of the momentum distribution and static structure factor for Coulombic systems. Unlike the unitary Fermi gas, we were unable to evaluate $\gamma(\p{q};\Lambda)$ in closed form.  Therefore, we turn to a perturbative calculation and expand the Q-space propagator 
\begin{eqnarray}
\frac{1}{\projQ_{\Lambda} H\projQ_{\Lambda}} &=& \frac{1}{\projQ_{\Lambda} T\projQ_{\Lambda}} -\frac{1}{\projQ_{\Lambda} T\projQ_{\Lambda}}V\frac{1}{\projQ_{\Lambda} T\projQ_{\Lambda}} +\frac{1}{\projQ_{\Lambda} T\projQ_{\Lambda}}V\frac{1}{\projQ_{\Lambda} T\projQ_{\Lambda}}V\frac{1}{\projQ_{\Lambda} T\projQ_{\Lambda}}+\ldots\,,\nonumber\\
\end{eqnarray}
which together with Eq.~\ref{eq:gammalambda} gives the first- and second-order contributions to $\gamma$
\begin{eqnarray}
\label{eq:gammaLO}
\gamma^{(1)}(\p{q};\Lambda) &=& -\int_{\Lambda}^{\infty}\frac{d^3q'}{(2\pi)^3}\,\frac{(2\pi)^3}{q^2}\delta^{3}(\p{q}-\p{q}')\frac{4\pi e^2}{q'^2}\nonumber\\
&=& -\frac{4\pi}{a_0q^4}\,,
\end{eqnarray}
and
\begin{eqnarray}
\gamma^{(2)}(\p{q};\Lambda) &=& -\int_{\Lambda}^{\infty}\frac{d^3q'}{(2\pi)^3}\frac{1}{q^2q'^2}\frac{4\pi e^2}{|\p{q}-\p{q}'|^2}\frac{4\pi e^2}{q'^2}\nonumber\\
&\approx& \frac{8}{a_0^2q^4\Lambda}\, 
\end{eqnarray}
where we've kept the leading term in $1/q$ for the second-order contribution and $a_0=\frac{\hbar^2}{e^2m}$ is the Bohr radius. We assume that perturbation theory is justified provided
\begin{equation}
\label{eq:pertvalidity}
\frac{\gamma^{(2)}}{\gamma^{(1)}} = \frac{2}{\pi}\frac{1}{a_0\Lambda} \ll1\quad\Rightarrow\quad \Lambda \gg \frac{2}{\pi}\frac{1}{a_0}\,,
\end{equation}
and restrict our attention to the leading term, Eq.~\ref{eq:gammaLO}. Inserting this into Eq.~\ref{eq:momdistfacA} gives
\begin{eqnarray}
\label{eq:momdistfaccoul}
n_{\p{q}}\approx \frac{16\pi^2}{q^8 a_0^2}Z^2_{\Lambda}\,\sum\limits_{\p{k,k',K}}\bra{\psilamalphaA}a^\dagger_{\frac{\p{K}}{2}+\p{k},\uparrow}a^\dagger_{\frac{\p{K}}{2}-\p{k},\downarrow}a_{\frac{\p{K}}{2}+\p{k'},\uparrow}a_{\frac{\p{K}}{2}-\p{k'},\downarrow}\ket{\psilamalphaA}\,.
\end{eqnarray}
Apart from the $Z_{\Lambda}$ factors and the evolved wave functions $\psilamalphaA$, this is very similar to the known result first derived by Kimball~\cite{1975JPhA....8.1513K}
\begin{equation}
\label{eq:momdistcoul}
n_{\p{q}}\approx \frac{16\pi^2}{q^8 a_0^2}\,\sum\limits_{\p{k,k',K}}\bra{\psiinfalphaA}a^\dagger_{\frac{\p{K}}{2}+\p{k},\uparrow}a^\dagger_{\frac{\p{K}}{2}-\p{k},\downarrow}a_{\frac{\p{K}}{2}+\p{k'},\uparrow}a_{\frac{\p{K}}{2}-\p{k'},\downarrow}\ket{\psiinfalphaA}\,.
\end{equation}
As with the unitary Fermi gas, one can make a heuristic argument that Eq.~\ref{eq:momdistfaccoul} and Eq.~\ref{eq:momdistcoul} are equivalent since $Z_{\Lambda}\rightarrow 1$ and $\psilamalphaA\rightarrow \psiinfalphaA$ as $\Lambda\rightarrow\infty$. More precisely, we will show in the Appendix that
\begin{eqnarray}
\label{eq:bareevolvedcontact}
\sum\limits_{\p{k,k',K}}\bra{\psiinfalphaA}a^\dagger_{\frac{\p{K}}{2}+\p{k},\uparrow}a^\dagger_{\frac{\p{K}}{2}-\p{k},\downarrow}a_{\frac{\p{K}}{2}+\p{k'},\uparrow}a_{\frac{\p{K}}{2}-\p{k'},\downarrow}\ket{\psiinfalphaA} \approx \nonumber \\
Z^2_{\Lambda}\left\{1 + \mathcal{O}\left(\frac{1}{ \Lambda a_0}\right)\right\}
\sum\limits_{\p{k,k',K}}\bra{\psilamalphaA}a^\dagger_{\frac{\p{K}}{2}+\p{k},\uparrow}a^\dagger_{\frac{\p{K}}{2}-\p{k},\downarrow}a_{\frac{\p{K}}{2}+\p{k'},\uparrow}a_{\frac{\p{K}}{2}-\p{k'},\downarrow}\ket{\psilamalphaA}\,,
\end{eqnarray}
so that Eqs.~\ref{eq:momdistfaccoul} and~\ref{eq:momdistcoul} are equivalent up to terms of order $\mathcal{O}(\frac{1}{\Lambda a_0})$, which are presumed to be small by virtue of Eq.~\ref{eq:pertvalidity}. 

\subsubsection{Static structure factor}
Turning next to the application of Eq.~\ref{eq:staticEff2} to Coulomb systems, our task is to evaluate the following term
\begin{equation}
\label{eq:Coulombintegral1}
\sum\limits_{\p{P},I}Z_{\Lambda}^2\,\gamma(\p{P+q};\Lambda)\gamma(\p{P};\Lambda)\,\,\rightarrow\,\,\left(\frac{4\pi Z_{\Lambda}}{a_0}\right)^2\,\int \frac{d^3P}{(2\pi)^3}\frac{\theta(|\p{P}|-\Lambda)\theta(|\p{P}+\p{q}|-\Lambda)}{|\p{P}+\p{q}|^4|\p{P}|^4}\,.
\end{equation}
As before, we split the integral into a sum of three terms 
\begin{equation}
I = \int \frac{d^3P}{(2\pi)^3}\frac{\theta(|\p{P}|-\Lambda)\theta(|\p{P}+\p{q}|-\Lambda)}{|\p{P}+\p{q}|^4|\p{P}|^4} = I_{\rm{high}} + I_{\rm{medium}} + I_{\rm{low}}\,
\end{equation}
where $I_{\rm{high}}$ corresponds to $|\p{P}| \ge \Lambda +|\p{q}|$, $I_{\rm{medium}}$ corresponds to $|\p{q}|-\Lambda \le |\p{P}| \le |\p{q}|+\Lambda$, and $I_{\rm{low}}$ corresponds $|\p{P}|\le |\p{q}| -\Lambda$. For $I_{\rm{high}}$ and $I_{\rm{low}}$, the angular integrals are trivial
\begin{eqnarray}
I_{\rm{high}} &=& \frac{1}{4\pi^2}\int_{\Lambda+q}^{\infty}P^2 dP\int_{-1}^{1}\frac{1}{P^4}\frac{1}{\left(P^2+q^2+2Pqx\right)^2} \nonumber\\
&\approx& \frac{1}{4\pi^2}\frac{1}{2\Lambda q^4}\,+\,\mathcal{O}\left(\frac{1}{q^5}\right)\,,\\
I_{\rm{low}}&=& \frac{1}{4\pi^2}\int_{\Lambda}^{q-\Lambda}P^2 dP\int_{-1}^{1}\frac{1}{P^4}\frac{1}{\left(P^2+q^2+2Pqx\right)^2} \nonumber\\
&\approx& \frac{1}{4\pi^2}\frac{5}{2\Lambda q^4} \,+\,\mathcal{O}\left(\frac{1}{q^5}\right)\,.
\end{eqnarray}
For $I_{\rm{medium}}$, the Heaviside theta functions restrict the angular integral
\begin{eqnarray}
\label{eq:Imedcoul}
I_{\rm{medium}} &=& \frac{1}{4\pi^2}\int_{q-\Lambda}^{q+\Lambda}P^2 dP\int_{x_{\rm{min}}}^{1}\frac{1}{P^4}\frac{1}{\left(P^2+q^2+2Pqx\right)^2} \nonumber\\
&\approx& \frac{1}{4\pi^2}\frac{1}{\Lambda q^4}\,+\,\mathcal{O}\left(\frac{1}{q^5}\right)\,,
\end{eqnarray}
where $x_{\rm{min}} = \frac{\Lambda^2-P^2-q^2}{2Pq}$. Combining Eqs.~\ref{eq:Coulombintegral1}-\ref{eq:Imedcoul} with Eq.~\ref{eq:staticEff2}, we find
\begin{eqnarray}
S_{\uparrow\!\downarrow}(\p{q}) &\approx&\frac{8\pi}{a_0}\frac{Z^2_{\Lambda}}{q^4}\left(1-\frac{2}{\pi}\frac{1}{\Lambda a_0}\right)\,\sum\limits_{\p{K,k,k'}}\, \bra{\psilamalphaA}a^\dagger_{\frac{\p{K}}{2}+\p{k},\uparrow}a^\dagger_{\frac{\p{K}}{2}-\p{k},\downarrow}a_{\frac{\p{K}}{2}+\p{k'},\downarrow}a_{\frac{\p{K}}{2}-\p{k'},\uparrow}\ket{\psilamalphaA}  \nonumber\\
&\approx& \frac{8\pi}{a_0}\frac{1}{q^4}\,\sum\limits_{\p{K,k,k'}}\, \bra{\psiinfalphaA}a^\dagger_{\frac{\p{K}}{2}+\p{k},\uparrow}a^\dagger_{\frac{\p{K}}{2}-\p{k},\downarrow}a_{\frac{\p{K}}{2}+\p{k'},\downarrow}a_{\frac{\p{K}}{2}-\p{k'},\uparrow}\ket{\psiinfalphaA}\,, 
\end{eqnarray}
where we've used Eq.~\ref{eq:pertvalidity} and Eq.~\ref{eq:bareevolvedcontact} to obtain the second line. Once again, this is in agreement with the previously known result of Kimball~\cite{1975JPhA....8.1513K}.

\section{Summary}  \label{sec:summary}
In this paper, we have used elementary RG arguments to show that, for general low-energy many-body states,  the high-momentum tails of momentum distributions and static structure factors factorize into the product of a universal function of momentum that is fixed (in leading-order) by two-body physics, and a state-dependent matrix element that is sensitive only to low-momentum structure of the many-body state, and is the same for both. This generalizes the results of Anderson \emph{et al.}, who derived analogous relations in the two-body system~\cite{Anderson:2010aq}, and suggests a possible interpretation of the universal high-momentum dependence and scaling behavior found in nuclear momentum distributions in the analysis of $(e,e'p)$ reactions. As a check, we have successfully applied our factorization relations to two well-studied systems, the unitary Fermi gas and the electron gas, reproducing known results for the high-momentum tails of each.

Our proof of factorization follows from decoupling and the separation of scales, and resembles aspects of the OPE in quantum field theory. Unfortunately, we have not been able to establish a precise connection. The main difference appears to be that, in a local quantum field theoretical  framework, the OPE offers a controlled expansion since the scaling dimension of a given local operator uniquely fixes the $\p{r}$-dependence of the corresponding Wilson coefficient, making the truncation of the expansion controllable.  In contrast, in the present paper we work in the domain of general non-relativistic quantum mechanics and do not require that the system is described by a local QFT. This relaxation of assumptions allows us to extend the notion of factorization and OPE-like methods to a wider class of problems, albeit in a less controlled fashion since we cannot make precise statements about the scaling properties of the operators kept/omitted in our expansions. Nevertheless, the methods presented in this paper may still be useful in low-energy nuclear physics, as they provide tools that allow us to parameterize the high-momentum components of operators that would normally require degrees of freedom that we do not retain. We can, for example, build effective operators containing state-independent functions of high momenta that can in principle be extracted from few-body data, and subsequently used to make predictions for high-momentum processes in $A$-body systems.

\begin{acknowledgments}
We thank Eric Anderson, Dick Furnstahl, Kai Hebeler, Heiko Hergert, Robert Perry, and Lucas Platter for useful comments and discussions.  This work was supported in part by the National Science Foundation under Grant Nos. PHY-0758125 and PHY-1068648. 

\end{acknowledgments}

\appendix
\section{Tan's contact at different resolutions }
In this appendix, we derive the following relation (Eq.~\ref{eq:contact} in the text) 
\begin{eqnarray}
\label{eq:contactapp}
Z_{\Lambda}^2 C(\Lambda) 
= C(\Lambda_0)\,,
\end{eqnarray}
where Tan's contact parameter is defined as
\begin{equation}
C(\Lambda_0) = g^2(\Lambda_0)\sum\limits_{\p{K,k,k'}}^{\Lambda_0}\, \bra{\psi^{\Lambda_0}_{\alpha,_A}}a^\dagger_{\frac{\p{K}}{2}+\p{k}}a^\dagger_{\frac{\p{K}}{2}-\p{k}}a_{\frac{\p{K}}{2}+\p{k'}}a_{\frac{\p{K}}{2}-\p{k'}}\ket{\psi^{\Lambda_0}_{\alpha,_A}}\,,
\end{equation}
and $g(\Lambda_0)$ is given by Eq.~\ref{Coupling}. Let us begin with the following relation 
\begin{eqnarray}
\label{eq:append1}
\sum\limits_{\p{K,k,k'}}^{\Lambda_0}\, \bra{\psi^{\Lambda_0}_{\alpha,_A}}a^\dagger_{\frac{\p{K}}{2}+\p{k}}a^\dagger_{\frac{\p{K}}{2}-\p{k}}a_{\frac{\p{K}}{2}+\p{k'}}a_{\frac{\p{K}}{2}-\p{k'}}\ket{\psi^{\Lambda_0}_{\alpha,_A}} &=&
\sum\limits_{\p{K,k,k'}}^{\Lambda_0}\,\bra{\psi^{\Lambda}_{\alpha,_A}}\left[a^\dagger_{\frac{\p{K}}{2}+\p{k}}a^\dagger_{\frac{\p{K}}{2}-\p{k}}\right]^{\Lambda}\,\left[a_{\frac{\p{K}}{2}+\p{k'}}a_{\frac{\p{K}}{2}-\p{k'}}\right]^{\Lambda}\ket{\psi^{\Lambda}_{\alpha,_A}}\,,\nonumber \\
\end{eqnarray}
where the evolved pair creation operators are given (within the same approximations made in the text) by

\begin{eqnarray}
\label{eq:paircreationevolved}
\left[a^\dagger_{\frac{\p{K}}{2}+\p{k}}a^\dagger_{\frac{\p{K}}{2}-\p{k}}\right]^{\Lambda} &=& a^\dagger_{\frac{\p{K}}{2}+\p{k}}a^\dagger_{\frac{\p{K}}{2}-\p{k}} +\sum_{\p{p}} C^{\Lambda}_{\p{k}}(\p{p},-\p{p})\,a^\dagger_{\frac{\p{K}}{2}+\p{p}}a^\dagger_{\frac{\p{K}}{2}-\p{p}} \,\nonumber\\
&\equiv& a^\dagger_{\frac{\p{K}}{2}+\p{k}}a^\dagger_{\frac{\p{K}}{2}-\p{k}} +
\contraction{\delta}{a^{\dagger}}{_{\frac{\p{K}}{2}+\p{k}}}{a}
\delta a^{\dagger}_{\frac{\p{K}}{2}+\p{k}}a^\dagger_{\frac{\p{K}}{2}-\p{k}}\,,
\end{eqnarray}
and similarly for the pair annihilation operators. Note that the form of Eq.~\ref{eq:paircreationevolved} is dictated by the two approximations made in the text, namely i) the neglect of higher-order corrections in Eq.~\ref{eq:adagger_lambda} for the evolved creation/annihilation operators and ii) the neglect of induced three- and higher-body operators in the evolution.  At this level of approximation, the right-hand side of Eq.~\ref{eq:append1} becomes 
\begin{equation}
\begin{aligned}
\label{eq:append2}
\rm{RHS\,\, of\,\,Eq.\,\,}\ref{eq:append1} = \sum\limits_{\p{K,k,k'}}^{\Lambda_0}&\,\Biggl[\bra{\psi^{\Lambda}_{\alpha,_A}}a^\dagger_{\frac{\p{K}}{2}+\p{k}}a^\dagger_{\frac{\p{K}}{2}-\p{k}}a_{\frac{\p{K}}{2}+\p{k'}}a_{\frac{\p{K}}{2}-\p{k'}}\ket{\psi^{\Lambda}_{\alpha,_A}}\\[.25 cm]
&+\,\,\bra{\psi^{\Lambda}_{\alpha,_A}}
\contraction{\delta}{a^\dagger}{_{\frac{\p{K}}{2}+\p{k}}}{a}
\delta a^\dagger_{\frac{\p{K}}{2}+\p{k}}a^\dagger_{\frac{\p{K}}{2}-\p{k}}
a_{\frac{\p{K}}{2}+\p{k'}}a_{\frac{\p{K}}{2}-\p{k'}}\ket{\psi^{\Lambda}_{\alpha,_A}} \\[.25 cm]
&+ \,\,
\bra{\psi^{\Lambda}_{\alpha,_A}}
a^\dagger_{\frac{\p{K}}{2}+\p{k}}a^\dagger_{\frac{\p{K}}{2}-\p{k}}
\contraction{}{a}{_{\frac{\p{K}}{2}+\p{k'}}\delta}{a}
a_{\frac{\p{K}}{2}+\p{k'}}\delta a_{\frac{\p{K}}{2}-\p{k'}}\ket{\psi^{\Lambda}_{\alpha,_A}} \\[.25 cm]
&+\,\, 
\bra{\psi^{\Lambda}_{\alpha,_A}}
\contraction{\delta}{a^\dagger}{_{\frac{\p{K}}{2}+\p{k}}}{a}
\delta a^\dagger_{\frac{\p{K}}{2}+\p{k}}a^\dagger_{\frac{\p{K}}{2}-\p{k}}
\contraction{}{a}{_{\frac{\p{K}}{2}+\p{k'}}\delta}{a}
a_{\frac{\p{K}}{2}+\p{k'}}\delta a_{\frac{\p{K}}{2}-\p{k'}}\ket{\psi^{\Lambda}_{\alpha,_A}}\Biggr]\,.
\end{aligned}
\end{equation}
Next, we split each summation into low- and high-momentum regions and use decoupling together with the asymptotic forms in Eqs.~\ref{eq:Cpp} and \ref{eq:Cqp} to simplify the resulting expressions. As a consequence of decoupling, the first term in Eq.~\ref{eq:append2} stays the same but with the summations effectively cutoff at $\Lambda \ll \Lambda_0$. The second term is given by 
\begin{eqnarray}
\sum\limits_{\p{K,p,k'}}^{\Lambda}\sum_{\p{k}}^{\Lambda_0}C^{\Lambda}_{\p{k}}(\p{p},-\p{p})\,\bra{\psi^{\Lambda}_{\alpha,_A}}a^\dagger_{\frac{\p{K}}{2}+\p{p}}a^\dagger_{\frac{\p{K}}{2}-\p{p}}a_{\frac{\p{K}}{2}+\p{k'}}a_{\frac{\p{K}}{2}-\p{k'}}\ket{\psi^{\Lambda}_{\alpha,_A}}\,,
\end{eqnarray}
where the sums over $\p{K},\p{p},$ and $\p{k}'$ are cutoff at $\Lambda$ due to decoupling. Splitting the sum over $\p{k}$ into low- and high-momentum regions and using Eqs.~\ref{eq:Cpp} and \ref{eq:Cqp} for $C^{\Lambda}_{\p{k}}(\p{p},-\p{p})$ gives
\begin{equation}
\begin{aligned}
\sum\limits_{\p{K,k,k'}}^{\Lambda_0}\bra{\psi^{\Lambda}_{\alpha,_A}}
\contraction{\delta}{a^\dagger}{_{\frac{\p{K}}{2}+\p{k}}}{a}
\delta a^\dagger_{\frac{\p{K}}{2}+\p{k}}a^\dagger_{\frac{\p{K}}{2}-\p{k}}
a_{\frac{\p{K}}{2}+\p{k'}}a_{\frac{\p{K}}{2}-\p{k'}}\ket{\psi^{\Lambda}_{\alpha,_A}} &\approx 
\\[.25 cm]
\bigl(Z_{\Lambda}-1 + Z_{\Lambda}\bar{\gamma}(\Lambda)\bigr)&\sum\limits_{\p{K,k,k'}}^{\Lambda}\, \bra{\psi^{\Lambda}_{\alpha,_A}}a^\dagger_{\frac{\p{K}}{2}+\p{k}}a^\dagger_{\frac{\p{K}}{2}-\p{k}}a_{\frac{\p{K}}{2}+\p{k'}}a_{\frac{\p{K}}{2}-\p{k'}}\ket{\psi^{\Lambda}_{\alpha,_A}} \,,
\end{aligned}
\end{equation}
where we've defined
\begin{equation}
\bar{\gamma}(\Lambda)\equiv\sum\limits_{\p{q}=\Lambda}^{\Lambda_0}\gamma(\p{q};\Lambda)\,.
\end{equation}
One easily finds that the third term in Eq.~\ref{eq:append2} gives the same contribution as the second.  The fourth and final term in Eq.~\ref{eq:append2} is given by
\begin{equation}
\begin{aligned}
\sum\limits_{\p{K,p,p'}}^{\Lambda}\sum\limits_{\p{k,k'}}^{\Lambda_0}C^{\Lambda}_{\p{k}}(\p{p},-\p{p})C^{\Lambda}_{\p{k'}}(\p{p'},-\p{p'})\,\bra{\psi^{\Lambda}_{\alpha,_A}}a^\dagger_{\frac{\p{K}}{2}+\p{p}}a^\dagger_{\frac{\p{K}}{2}-\p{p}}a_{\frac{\p{K}}{2}+\p{p'}}&a_{\frac{\p{K}}{2}-\p{p'}}\ket{\psi^{\Lambda}_{\alpha,_A}}\,.
\end{aligned}
\end{equation}
As before, decoupling implies the sums over $\p{K},\p{p}$, and $\p{p'}$ are cutoff at $\Lambda$.
Splitting the unrestricted sums over $\p{k}$ and $\p{k'}$ into low- and high-momentum regions then gives
\begin{equation}
\begin{aligned}
\sum\limits_{\p{K,k,k'}}^{\Lambda_0}
\bra{\psi^{\Lambda}_{\alpha,_A}}
\contraction{\delta}{a^\dagger}{_{\frac{\p{K}}{2}+\p{k}}}{a}
\delta a^\dagger_{\frac{\p{K}}{2}+\p{k}}a^\dagger_{\frac{\p{K}}{2}-\p{k}}
\contraction{}{a}{_{\frac{\p{K}}{2}+\p{k'}}\delta}{a}
a_{\frac{\p{K}}{2}+\p{k'}}\delta a_{\frac{\p{K}}{2}-\p{k'}}\ket{\psi^{\Lambda}_{\alpha,_A}} &\approx
\biggl((Z_{\Lambda}-1)^2 + 2Z_{\Lambda}(Z_{\Lambda}-1)\bar{\gamma}(\Lambda) + Z^2_{\Lambda}\bar{\gamma}^2(\Lambda)\biggr)\\
&\times\,\,\sum\limits_{\p{K,k,k'}}^{\Lambda}\, \bra{\psi^{\Lambda}_{\alpha,_A}}a^\dagger_{\frac{\p{K}}{2}+\p{k}}a^\dagger_{\frac{\p{K}}{2}-\p{k}}a_{\frac{\p{K}}{2}+\p{k'}}a_{\frac{\p{K}}{2}-\p{k'}}\ket{\psi^{\Lambda}_{\alpha,_A}} \,.
\end{aligned}
\end{equation}
Collecting terms and simplifying, Eq.~\ref{eq:append1} becomes
\begin{equation}
\begin{aligned}
\label{eq:append3}
\sum\limits_{\p{K,k,k'}}^{\Lambda_0}\, \bra{\psi^{\Lambda_0}_{\alpha,_A}}a^\dagger_{\frac{\p{K}}{2}+\p{k}}a^\dagger_{\frac{\p{K}}{2}-\p{k}}a_{\frac{\p{K}}{2}+\p{k'}}a_{\frac{\p{K}}{2}-\p{k'}}\ket{\psi^{\Lambda_0}_{\alpha,_A}} &\approx\, Z^2_{\Lambda}\left(\bar{\gamma}(\Lambda) + 1\right)^2\\
&\times\,\,
\sum\limits_{\p{K,k,k'}}^{\Lambda}\, \bra{\psi^{\Lambda}_{\alpha,_A}}a^\dagger_{\frac{\p{K}}{2}+\p{k}}a^\dagger_{\frac{\p{K}}{2}-\p{k}}a_{\frac{\p{K}}{2}+\p{k'}}a_{\frac{\p{K}}{2}-\p{k'}}\ket{\psi^{\Lambda}_{\alpha,_A}} \,.
\end{aligned}
\end{equation} 

For the unitary Fermi gas, we can evaluate $\bar{\gamma}$ analytically using Eq.~\ref{eq:gammaUFG}
\begin{equation}
\bar{\gamma}(\Lambda) = \sum\limits_{\p{q}=\Lambda}^{\Lambda_0}\gamma(\p{q};\Lambda)\quad \rightarrow\quad \int_{\Lambda}^{\Lambda_0}\frac{d^3\p{q}}{(2\pi)^3}\frac{-g(\Lambda)}{q^2}\, =\, -\frac{g(\Lambda)}{2\pi^2}\,\left(\Lambda_0 - \Lambda\right) \,.
\end{equation}
Inserting this into Eq.~\ref{eq:append3} and using the identity
\begin{equation}
\frac{g(\Lambda)}{g(\Lambda_0)} = 1 - \frac{g(\Lambda)}{2\pi^2}\,\left(\Lambda_0 - \Lambda\right)\,,
\end{equation}
one finds the desired relation
\begin{equation}
\begin{aligned}
C(\Lambda_0)&=g^2(\Lambda_0)\sum\limits_{\p{K,k,k'}}^{\Lambda_0}\, \bra{\psi^{\Lambda_0}_{\alpha,_A}}a^\dagger_{\frac{\p{K}}{2}+\p{k}}a^\dagger_{\frac{\p{K}}{2}-\p{k}}a_{\frac{\p{K}}{2}+\p{k'}}a_{\frac{\p{K}}{2}-\p{k'}}\ket{\psi^{\Lambda_0}_{\alpha,_A}}\\
&\approx Z^2_{\Lambda}g^2(\Lambda)\sum\limits_{\p{K,k,k'}}^{\Lambda}\, \bra{\psi^{\Lambda}_{\alpha,_A}}a^\dagger_{\frac{\p{K}}{2}+\p{k}}a^\dagger_{\frac{\p{K}}{2}-\p{k}}a_{\frac{\p{K}}{2}+\p{k'}}a_{\frac{\p{K}}{2}-\p{k'}}\ket{\psi^{\Lambda}_{\alpha,_A}}\\
&= Z^2_{\Lambda}C(\Lambda)\,.
\end{aligned}
\end{equation}

\section{Coulomb pair-distribution function at different resolutions}
In this appendix, we derive the following relation (Eq.~\ref{eq:bareevolvedcontact} in the text) for the Coulomb gas,
\begin{equation}
\begin{aligned}
\label{eq:bareevolvedcontactapp}
\sum\limits_{\p{k,k',K}}\bra{\psiinfalphaA}a^\dagger_{\frac{\p{K}}{2}+\p{k}}a^\dagger_{\frac{\p{K}}{2}-\p{k}}a_{\frac{\p{K}}{2}+\p{k'}}a_{\frac{\p{K}}{2}-\p{k'}}\ket{\psiinfalphaA} &\approx  \\
Z^2_{\Lambda}\left\{1 + \mathcal{O}\left(\frac{1}{ \Lambda a_0}\right)\right\}&
\sum\limits_{\p{k,k',K}}\bra{\psilamalphaA}a^\dagger_{\frac{\p{K}}{2}+\p{k}}a^\dagger_{\frac{\p{K}}{2}-\p{k}}a_{\frac{\p{K}}{2}+\p{k'}}a_{\frac{\p{K}}{2}-\p{k'}}\ket{\psilamalphaA}\,.
\end{aligned}
\end{equation}
Since the proof closely mirrors what was done for the unitary Fermi gas, we begin from Eq.~\ref{eq:append3} with $\Lambda_0\rightarrow \infty$
\begin{equation}
\begin{aligned}
\label{eq:append4}
\sum\limits_{\p{K,k,k'}}^{\infty}\, \bra{\psi^{\infty}_{\alpha,_A}}a^\dagger_{\frac{\p{K}}{2}+\p{k}}a^\dagger_{\frac{\p{K}}{2}-\p{k}}a_{\frac{\p{K}}{2}+\p{k'}}a_{\frac{\p{K}}{2}-\p{k'}}\ket{\psi^{\infty}_{\alpha,_A}} &\approx\, Z^2_{\Lambda}\left(\bar{\gamma}(\Lambda) + 1\right)^2\\
&\times\,\,
\sum\limits_{\p{K,k,k'}}^{\Lambda}\, \bra{\psi^{\Lambda}_{\alpha,_A}}a^\dagger_{\frac{\p{K}}{2}+\p{k}}a^\dagger_{\frac{\p{K}}{2}-\p{k}}a_{\frac{\p{K}}{2}+\p{k'}}a_{\frac{\p{K}}{2}-\p{k'}}\ket{\psi^{\Lambda}_{\alpha,_A}} \,.
\end{aligned}
\end{equation} 
Using the leading perturbative expression for $\gamma(\p{q};\Lambda)$ in Eq.~\ref{eq:gammaLO}, we have
\begin{equation}
\bar{\gamma}(\Lambda) = \sum_{\p{q}=\Lambda}^{\infty} \gamma(\p{q};\Lambda)\quad\rightarrow\quad \int_{\Lambda}^{\infty}\frac{d^3\p{q}}{(2\pi)^3} \frac{-4\pi}{a_0}\frac{1}{q^4} = -\frac{2}{\pi}\frac{1}{\Lambda a_0}\,.
\end{equation}
Inserting this into Eq.~\ref{eq:append4} then gives 
\begin{equation}
\begin{aligned}
\label{eq:append5}
\sum\limits_{\p{K,k,k'}}^{\infty}\, \bra{\psi^{\infty}_{\alpha,_A}}a^\dagger_{\frac{\p{K}}{2}+\p{k}}a^\dagger_{\frac{\p{K}}{2}-\p{k}}a_{\frac{\p{K}}{2}+\p{k'}}a_{\frac{\p{K}}{2}-\p{k'}}\ket{\psi^{\infty}_{\alpha,_A}} &\approx\, Z^2_{\Lambda}\left(1-\frac{2}{\pi}\frac{1}{\Lambda a_0}\right)^2\\
&\times\,\,
\sum\limits_{\p{K,k,k'}}^{\Lambda}\, \bra{\psi^{\Lambda}_{\alpha,_A}}a^\dagger_{\frac{\p{K}}{2}+\p{k}}a^\dagger_{\frac{\p{K}}{2}-\p{k}}a_{\frac{\p{K}}{2}+\p{k'}}a_{\frac{\p{K}}{2}-\p{k'}}\ket{\psi^{\Lambda}_{\alpha,_A}} \,,
\end{aligned}
\end{equation} 
which is equivalent to Eq.~\ref{eq:bareevolvedcontactapp} since $\Lambda a_0 \gg 1$ by assumption.

\bibliographystyle{apsrev}
\bibliography{vlowk_refs}

\begin{thebibliography}{20}
\expandafter\ifx\csname natexlab\endcsname\relax\def\natexlab#1{#1}\fi
\expandafter\ifx\csname bibnamefont\endcsname\relax
  \def\bibnamefont#1{#1}\fi
\expandafter\ifx\csname bibfnamefont\endcsname\relax
  \def\bibfnamefont#1{#1}\fi
\expandafter\ifx\csname citenamefont\endcsname\relax
  \def\citenamefont#1{#1}\fi
\expandafter\ifx\csname url\endcsname\relax
  \def\url#1{\texttt{#1}}\fi
\expandafter\ifx\csname urlprefix\endcsname\relax\def\urlprefix{URL }\fi
\providecommand{\bibinfo}[2]{#2}
\providecommand{\eprint}[2][]{\url{#2}}

\bibitem[{\citenamefont{Anderson et~al.}(2010)\citenamefont{Anderson, Bogner,
  Furnstahl, and Perry}}]{Anderson:2010aq}
\bibinfo{author}{\bibfnamefont{E.}~\bibnamefont{Anderson}},
  \bibinfo{author}{\bibfnamefont{S.}~\bibnamefont{Bogner}},
  \bibinfo{author}{\bibfnamefont{R.}~\bibnamefont{Furnstahl}},
  \bibnamefont{and} \bibinfo{author}{\bibfnamefont{R.}~\bibnamefont{Perry}},
  \bibinfo{journal}{Phys.Rev.} \textbf{\bibinfo{volume}{C82}},
  \bibinfo{pages}{054001} (\bibinfo{year}{2010}), \eprint{1008.1569}.

\bibitem[{\citenamefont{Bogner et~al.}(2003{\natexlab{a}})\citenamefont{Bogner,
  Kuo, and Schwenk}}]{Bogner:2003wn}
\bibinfo{author}{\bibfnamefont{S.~K.} \bibnamefont{Bogner}},
  \bibinfo{author}{\bibfnamefont{T.~T.~S.} \bibnamefont{Kuo}},
  \bibnamefont{and} \bibinfo{author}{\bibfnamefont{A.}~\bibnamefont{Schwenk}},
  \bibinfo{journal}{Phys. Rept.} \textbf{\bibinfo{volume}{386}},
  \bibinfo{pages}{1} (\bibinfo{year}{2003}{\natexlab{a}}),
  \eprint{nucl-th/0305035}.

\bibitem[{\citenamefont{Bogner et~al.}(2010)\citenamefont{Bogner, Furnstahl,
  and Schwenk}}]{Bogner:2009bt}
\bibinfo{author}{\bibfnamefont{S.~K.} \bibnamefont{Bogner}},
  \bibinfo{author}{\bibfnamefont{R.~J.} \bibnamefont{Furnstahl}},
  \bibnamefont{and} \bibinfo{author}{\bibfnamefont{A.}~\bibnamefont{Schwenk}},
  \bibinfo{journal}{Prog. Part. Nucl. Phys.} \textbf{\bibinfo{volume}{65}},
  \bibinfo{pages}{94} (\bibinfo{year}{2010}), \eprint{0912.3688}.

\bibitem[{\citenamefont{Furnstahl}(2012)}]{Furnstahl:2012fn}
\bibinfo{author}{\bibfnamefont{R.}~\bibnamefont{Furnstahl}}
  (\bibinfo{year}{2012}), \eprint{1203.1779}.

\bibitem[{\citenamefont{Epelbaum et~al.}(1999)\citenamefont{Epelbaum,
  Gl{\"o}ckle, Kruger, and Meissner}}]{Epelbaum:1998na}
\bibinfo{author}{\bibfnamefont{E.}~\bibnamefont{Epelbaum}},
  \bibinfo{author}{\bibfnamefont{W.}~\bibnamefont{Gl{\"o}ckle}},
  \bibinfo{author}{\bibfnamefont{A.}~\bibnamefont{Kruger}}, \bibnamefont{and}
  \bibinfo{author}{\bibfnamefont{U.-G.} \bibnamefont{Meissner}},
  \bibinfo{journal}{Nucl. Phys. A} \textbf{\bibinfo{volume}{645}},
  \bibinfo{pages}{413} (\bibinfo{year}{1999}), \eprint{nucl-th/9809084}.

\bibitem[{\citenamefont{Bogner et~al.}(2003{\natexlab{b}})\citenamefont{Bogner,
  Kuo, Schwenk, Entem, and Machleidt}}]{Bogner:2001gq}
\bibinfo{author}{\bibfnamefont{S.~K.} \bibnamefont{Bogner}},
  \bibinfo{author}{\bibfnamefont{T.~T.~S.} \bibnamefont{Kuo}},
  \bibinfo{author}{\bibfnamefont{A.}~\bibnamefont{Schwenk}},
  \bibinfo{author}{\bibfnamefont{D.~R.} \bibnamefont{Entem}}, \bibnamefont{and}
  \bibinfo{author}{\bibfnamefont{R.}~\bibnamefont{Machleidt}},
  \bibinfo{journal}{Phys. Lett. B} \textbf{\bibinfo{volume}{576}},
  \bibinfo{pages}{265} (\bibinfo{year}{2003}{\natexlab{b}}),
  \eprint{nucl-th/0108041}.

\bibitem[{\citenamefont{Bogner et~al.}(2007{\natexlab{a}})\citenamefont{Bogner,
  Furnstahl, Ramanan, and Schwenk}}]{Bogner:2006vp}
\bibinfo{author}{\bibfnamefont{S.~K.} \bibnamefont{Bogner}},
  \bibinfo{author}{\bibfnamefont{R.~J.} \bibnamefont{Furnstahl}},
  \bibinfo{author}{\bibfnamefont{S.}~\bibnamefont{Ramanan}}, \bibnamefont{and}
  \bibinfo{author}{\bibfnamefont{A.}~\bibnamefont{Schwenk}},
  \bibinfo{journal}{Nucl. Phys. A} \textbf{\bibinfo{volume}{784}},
  \bibinfo{pages}{79} (\bibinfo{year}{2007}{\natexlab{a}}),
  \eprint{nucl-th/0609003}.

\bibitem[{\citenamefont{Bogner et~al.}(2007{\natexlab{b}})\citenamefont{Bogner,
  Furnstahl, and Perry}}]{Bogner:2006pc}
\bibinfo{author}{\bibfnamefont{S.~K.} \bibnamefont{Bogner}},
  \bibinfo{author}{\bibfnamefont{R.~J.} \bibnamefont{Furnstahl}},
  \bibnamefont{and} \bibinfo{author}{\bibfnamefont{R.~J.} \bibnamefont{Perry}},
  \bibinfo{journal}{Phys. Rev. C} \textbf{\bibinfo{volume}{75}},
  \bibinfo{pages}{061001} (\bibinfo{year}{2007}{\natexlab{b}}),
  \eprint{nucl-th/0611045}.

\bibitem[{\citenamefont{Jurgenson et~al.}(2008)\citenamefont{Jurgenson, Bogner,
  Furnstahl, and Perry}}]{Jurgenson:2007td}
\bibinfo{author}{\bibfnamefont{E.~D.} \bibnamefont{Jurgenson}},
  \bibinfo{author}{\bibfnamefont{S.~K.} \bibnamefont{Bogner}},
  \bibinfo{author}{\bibfnamefont{R.~J.} \bibnamefont{Furnstahl}},
  \bibnamefont{and} \bibinfo{author}{\bibfnamefont{R.~J.} \bibnamefont{Perry}},
  \bibinfo{journal}{Phys. Rev. C} \textbf{\bibinfo{volume}{78}},
  \bibinfo{pages}{014003} (\bibinfo{year}{2008}), \eprint{0711.4252}.

\bibitem[{\citenamefont{Frankfurt
  et~al.}(2008{\natexlab{a}})\citenamefont{Frankfurt, Sargsian, and
  Strikman}}]{Frankfurt:2008zv}
\bibinfo{author}{\bibfnamefont{L.}~\bibnamefont{Frankfurt}},
  \bibinfo{author}{\bibfnamefont{M.}~\bibnamefont{Sargsian}}, \bibnamefont{and}
  \bibinfo{author}{\bibfnamefont{M.}~\bibnamefont{Strikman}},
  \bibinfo{journal}{Int. J. Mod. Phys. A} \textbf{\bibinfo{volume}{23}},
  \bibinfo{pages}{2991} (\bibinfo{year}{2008}{\natexlab{a}}),
  \eprint{0806.4412}.

\bibitem[{\citenamefont{Pieper et~al.}(1992)\citenamefont{Pieper, Wiringa, and
  Pandharipande}}]{Pieper:1992gr}
\bibinfo{author}{\bibfnamefont{S.~C.} \bibnamefont{Pieper}},
  \bibinfo{author}{\bibfnamefont{R.~B.} \bibnamefont{Wiringa}},
  \bibnamefont{and} \bibinfo{author}{\bibfnamefont{V.~R.}
  \bibnamefont{Pandharipande}}, \bibinfo{journal}{Phys. Rev.}
  \textbf{\bibinfo{volume}{C46}}, \bibinfo{pages}{1741} (\bibinfo{year}{1992}).

\bibitem[{\citenamefont{Wilson}(1969)}]{Wilson:1969zs}
\bibinfo{author}{\bibfnamefont{K.~G.} \bibnamefont{Wilson}},
  \bibinfo{journal}{Phys.Rev.} \textbf{\bibinfo{volume}{179}},
  \bibinfo{pages}{1499} (\bibinfo{year}{1969}).

\bibitem[{\citenamefont{Wilson and Zimmermann}(1972)}]{Wilson:1972ee}
\bibinfo{author}{\bibfnamefont{K.}~\bibnamefont{Wilson}} \bibnamefont{and}
  \bibinfo{author}{\bibfnamefont{W.}~\bibnamefont{Zimmermann}},
  \bibinfo{journal}{Commun.Math.Phys.} \textbf{\bibinfo{volume}{24}},
  \bibinfo{pages}{87} (\bibinfo{year}{1972}).

\bibitem[{Lep(1997)}]{Lepage:1997cs}
\emph{\bibinfo{title}{{G.P. Lepage, ``How to Renormalize the Schr{\"o}dinger
  Equation'', Lectures given at 9th Jorge Andre Swieca Summer School: Particles
  and Fields, Sao Paulo, Brazil, February, 1997, nucl-th/9706029}}}
  (\bibinfo{year}{1997}).

\bibitem[{\citenamefont{Felline et~al.}(2003)\citenamefont{Felline, Mehta,
  Piekarewicz, and Shepard}}]{Felline:2003mi}
\bibinfo{author}{\bibfnamefont{C.}~\bibnamefont{Felline}},
  \bibinfo{author}{\bibfnamefont{N.}~\bibnamefont{Mehta}},
  \bibinfo{author}{\bibfnamefont{J.}~\bibnamefont{Piekarewicz}},
  \bibnamefont{and} \bibinfo{author}{\bibfnamefont{J.}~\bibnamefont{Shepard}},
  \bibinfo{journal}{Phys.Rev.} \textbf{\bibinfo{volume}{C68}},
  \bibinfo{pages}{034003} (\bibinfo{year}{2003}), \eprint{nucl-th/0305007}.

\bibitem[{\citenamefont{Frankfurt
  et~al.}(2008{\natexlab{b}})\citenamefont{Frankfurt, Sargsian, and
  Strikman}}]{Frankfurt:2009vv}
\bibinfo{author}{\bibfnamefont{L.}~\bibnamefont{Frankfurt}},
  \bibinfo{author}{\bibfnamefont{M.}~\bibnamefont{Sargsian}}, \bibnamefont{and}
  \bibinfo{author}{\bibfnamefont{M.}~\bibnamefont{Strikman}},
  \bibinfo{journal}{AIP Conf. Proc.} \textbf{\bibinfo{volume}{1056}},
  \bibinfo{pages}{322} (\bibinfo{year}{2008}{\natexlab{b}}),
  \eprint{0901.2340}.

\bibitem[{\citenamefont{Braaten and Platter}(2008)}]{Braaten:2008uh}
\bibinfo{author}{\bibfnamefont{E.}~\bibnamefont{Braaten}} \bibnamefont{and}
  \bibinfo{author}{\bibfnamefont{L.}~\bibnamefont{Platter}},
  \bibinfo{journal}{Phys.Rev.Lett.} \textbf{\bibinfo{volume}{100}},
  \bibinfo{pages}{205301} (\bibinfo{year}{2008}), \eprint{0803.1125}.

\bibitem[{\citenamefont{{Tan}}(2008)}]{2008AnPhy.323.2971T}
\bibinfo{author}{\bibfnamefont{S.}~\bibnamefont{{Tan}}},
  \bibinfo{journal}{Annals of Physics} \textbf{\bibinfo{volume}{323}},
  \bibinfo{pages}{2971} (\bibinfo{year}{2008}),
  \eprint{arXiv:cond-mat/0508320}.

\bibitem[{\citenamefont{Braaten}(2012)}]{Braaten:2010if}
\bibinfo{author}{\bibfnamefont{E.}~\bibnamefont{Braaten}},
  \bibinfo{journal}{Lect.Notes Phys.} \textbf{\bibinfo{volume}{836}},
  \bibinfo{pages}{193} (\bibinfo{year}{2012}), \eprint{1008.2922}.

\bibitem[{\citenamefont{{Kimball}}(1975)}]{1975JPhA....8.1513K}
\bibinfo{author}{\bibfnamefont{J.~C.} \bibnamefont{{Kimball}}},
  \bibinfo{journal}{Journal of Physics A Mathematical General}
  \textbf{\bibinfo{volume}{8}}, \bibinfo{pages}{1513} (\bibinfo{year}{1975}).

\end{thebibliography}

\end{document}